\documentclass[]{pasj02} 
\usepackage[switch,mathlines]{lineno} 
\usepackage{natbib,bm} 

\jyear{2026}

\newcommand{\beq}{\begin{equation}}
\newcommand{\eeq}{\end{equation}}
\newcommand{\beqn}{\begin{eqnarray}}
\newcommand{\eeqn}{\end{eqnarray}}

\newcommand{\dint}{\displaystyle\int}
\newcommand{\pfrac}[2]{ \biggl(\dfrac{#1}{#2}\biggr) }

\begin{document}

\title{Bridging the gap: consistent modeling of protoplanetary disk heating and gap formation by planet-induced spiral shocks}

\author{
Satoshi \textsc{Okuzumi}\altaffilmark{1}\altemailmark\orcid{0000-0002-1886-0880}\email{okuzumi@eps.sci.isct.ac.jp},
Takayuki \textsc{Muto}\altaffilmark{2},
Ryosuke T. \textsc{Tominaga}\altaffilmark{1},
Shizu \textsc{Shimizu}\altaffilmark{1}
}

\altaffiltext{1}{Department of Earth and Planetary Sciences, Institute of Science Tokyo, 2-12-1 Ookayama, Meguro, Tokyo 152-8551, Japan}
\altaffiltext{2}{Division of Liberal Arts, Kogakuin University, 1-24-2 Nishi-Shinjuku, Shinjuku, Tokyo 163-8677, Japan}


\KeyWords{hydrodynamics --- planet--disk interactions ---  planets and satellites: formation --- protoplanetary disks} 

\maketitle

\begin{abstract}
A giant planet embedded in a protoplanetary disk excites spiral density waves, which steepen into shocks as they propagate away from the planet. These shocks lead to secular disk heating and gap opening, both of which can have important implications for the evolution of solids near the planet. To date, these two effects have largely been modeled independently. In this study, we present a self-consistent model that unifies these processes by linking shock heating and angular momentum deposition through the entropy jumps across the spiral shocks. We show that this model accurately reproduces the temperature and surface density profiles around the planet's orbit, as obtained from two-dimensional hydrodynamic simulations with standard $\alpha$ viscosity and $\beta$ thermal relaxation prescriptions. Furthermore, by incorporating an empirically derived scaling law for the radial distribution of the entropy jump, we construct a fully analytic model that self-consistently predicts the temperature and surface density structures of disks hosting a giant planet. This work represents a first step toward understanding how {a giant planet forming in the inner disk region influences the distribution and composition} of second-generation planets and planetesimals in its vicinity.

\end{abstract}


\section{Introduction}
\label{sec:intro}
Planets form in dusty gas disks surrounding young stars. The classical core-accretion model of planet formation \citep[e.g.,][]{Mizuno80,Pollack96} assumes that solid planetary embryos form from a subdisk of kilometer-sized planetesimals. However, multiple recent lines of evidence suggest that giant planets can form while protoplanetary gas disks still contain abundant solids required for the formation of smaller planets and planetesimals. High-resolution radio observations have revealed rich substructures, including rings, gaps, arcs, and spirals, in the gas and dust distributions of these disks \citep[for reviews, see][]{Andrews20,Bae23}, possibly indicating ongoing giant planet formation in still-dusty disks. Additionally, the nucleosynthetic isotopic dichotomy observed in solar system solids \citep[for reviews, see][]{Kleine20,Kruijer20} suggests that solid bodies in the solar system may have formed in two spatially separated regions of the solar nebula. One proposed candidate for the barrier separating these regions is proto-Jupiter that had already formed in the nebula \citep[e.g.,][]{Kruijer17}. These findings highlight the need to understand how early-formed giant planets influence the distribution, composition, and evolution of the gas and dust remaining in disks.

\begin{figure*}
\begin{center}
\includegraphics[width=0.75\hsize, bb=0 0 1600 780]{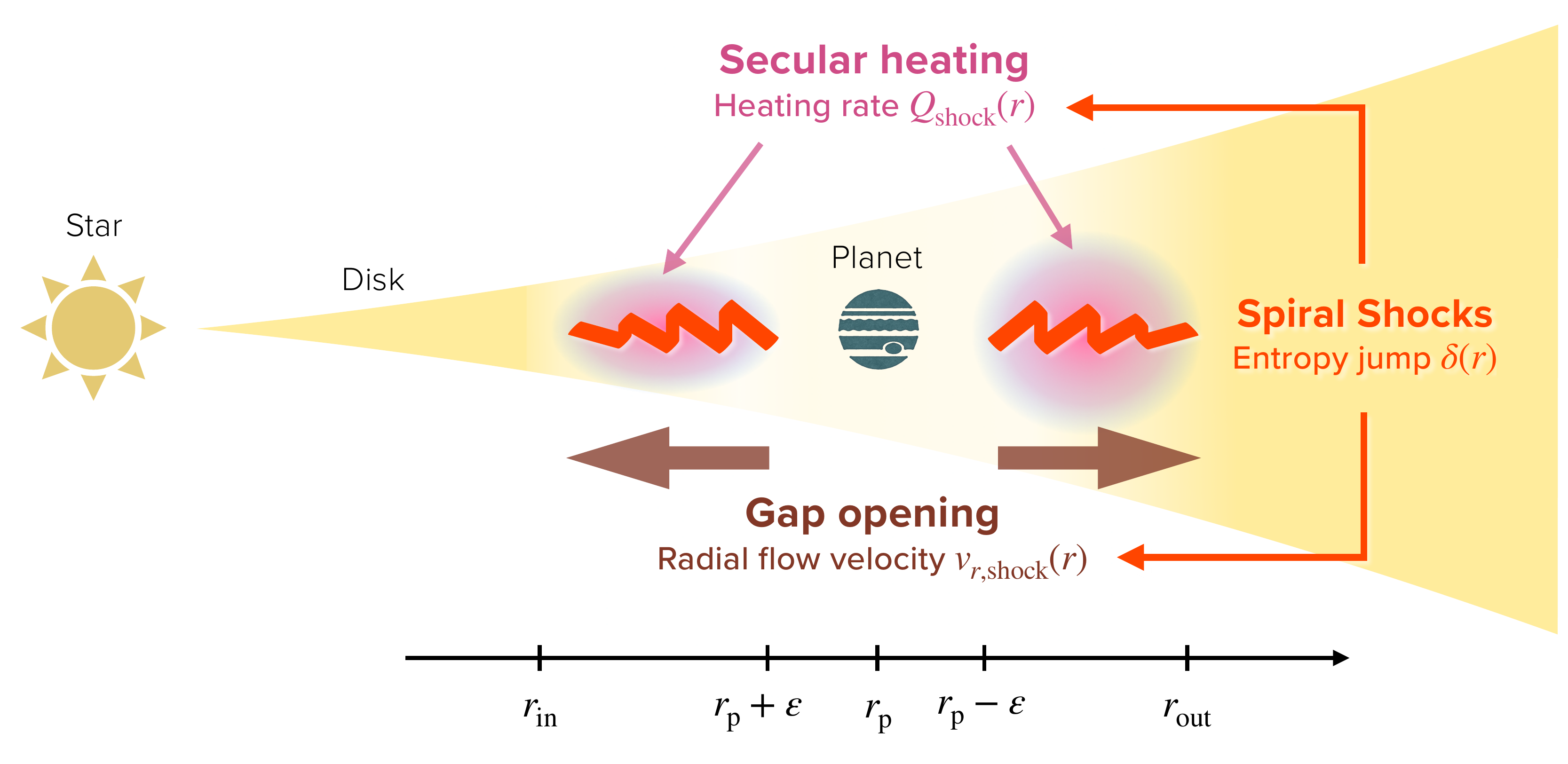}
\end{center}
\caption{Schematic illustration of the dual role of planet-induced spiral shocks \citep[see also][]{Rafikov16}. Entropy generation at the shock fronts leads to secular, irreversible heating of the disk, while angular momentum deposition by the dissipating shocks drives gap opening. The entropy jumps across the shocks, $\delta$, determine the shock heating rate $Q_{\rm shock}$ (equation~\eqref{eq:Qshock}; see also \citealt{Ono25}).
The connection between shock-induced heating and angular momentum transfer (equation~\eqref{eq:Lambda_shock}, \citealt{GoodmanRafikov01,Rafikov16,ArzamasskiyRafikov18}) also relates $\delta$ to the gap-opening flow velocity $v_{r,\rm shock}$ 
(equation~\eqref{eq:vr_shock}). 
}
\label{fig:concept}
\end{figure*}
Planet-induced density waves play a crucial role in planet--disk interactions (for a recent review, see \citealt{Paardekooper23}). The gravity of a planet drives spiral waves in a differentially rotating disk \citep{Goldreich79}.  
Due to the fluid's nonlinearity, these waves inevitably steepen into shocks as they travel away from the planet \citep{GoodmanRafikov01,Rafikov02a}. The dissipation of these spiral shocks affects the disk structure in two main ways (figure \ref{fig:concept}). First, entropy generation at the shock fronts leads to irreversible, secular heating of the disk material \citep{Richert15,Lyra16,Rafikov16,Ono25}\footnote{As noted by \citet{Rafikov16}, this secular heating is distinct from the heating and cooling associated with the adiabatic compression and expansion by the spiral waves (not necessarily shocks), which occur periodically and do not result in secular heating.}. 
Shock-mediated heating by giant planets can substantially {perturb the temperature of} {the disk's inner ($\lesssim$ 10 au), optically thick region} and even shift the snow line, where water ice sublimates \citep{Ziampras20}.
This effect could be particularly important in disks whose accretion is driven by magnetic winds, as wind-driven accretion tends to take place near the disk surface and may not contribute to midplane heating \citep{Mori19,Mori21,Bethune20,Kondo23}. 
Second, the dissipating spiral shocks transfer their angular momentum onto the fluid, resulting in the opening of a gap in the disk \citep[e.g.,][]{Rafikov02b,KanagawaTanaka15,KanagawaTanaka17,Duffell15}. 
The outer edge of the gas gap acts as a trap for inward-drifting dust grains \citep{Whipple72,Rice06,Paardekooper06}, with significant implications for subsequent planetesimal and planet formation \citep[e.g.,][]{Lyra09,Ayliffe12,Kobayashi12,Chatterjee14,Zhu14,Stammler19,Eriksson20,Shibaike20}, radial compositional gradients within the disk \citep[e.g.,][]{Morbidelli16,Desch18,Kalyaan21,Homma24}, and the disk's observational appearance  \citep[e.g.,][]{PinillaBenisty12,ZhuNelson12,Dullemond18}.

The aforementioned effects of planet-induced spiral shocks, namely secular heating and gap opening, are often modeled separately \citep[e.g.,][]{Duffell13,Duffell15,FungShi14,KanagawaMuto15,KanagawaTanaka15,KanagawaTanaka17,Ono25}. However, since both arise from the dissipation of the same spiral waves, they can, in principle, be modeled in a self-consistent manner. Indeed, it has been predicted  \citep{GoodmanRafikov01,Rafikov16} and subsequently confirmed  \citep{ArzamasskiyRafikov18} that the heating rate and angular momentum deposition rate associated with spiral shocks are mutually related. Therefore, given one of the two, the other can be inferred.  Recently, \citet{Ono25} numerically found that the specific entropy jumps across spiral shocks, which determine the shock heating rate \citep{Rafikov16}, follow common scaling relations with respect to orbital radius, planetary mass, and viscosity. Notably, they found that the specific entropy jumps are explicitly independent of the radial surface density profile. In this study, we utilize this property of the entropy jumps to {develop self-consistent models for the temperature and surface density structure around planet-induced gaps in the inner, slowly cooling region of protoplanetary disks}. 

The outline of this paper is as follows. In section~\ref{sec:simulations}, we present two-dimensional (2D) hydrodynamical simulations of protoplanetary disks with an embedded planet to provide the radial profiles of the disk surface density and temperature, as well as the entropy jumps across planet-induced spiral shocks, which are used in the subsequent analysis. In section~\ref{sec:bridge}, we use the relation between the shock heating rate and angular momentum deposition rate to derive the shock-induced radial gas velocity as a function of the entropy jump. We also demonstrate that the balance between the shock-induced and viscosity-driven flows is a key characteristic of a steady-state planet-induced gap.
In section~\ref{sec:semi}, we utilize the expression for the shock-induced radial velocity to predict the disk temperature and surface density profiles in the planet-carved gap from the radial profile of spiral shocks' entropy jumps. 
In section~\ref{sec:analytic}, we combine the results presented in sections~\ref{sec:simulations} and \ref{sec:semi} to derive fully analytic, self-consistent expressions for the gap temperature and surface density profiles. We discuss limitations of our results and directions for future work in section~\ref{sec:discussion}, and summarize key conclusions in section~\ref{sec:conclusions}.

\section{Simulation} 
\label{sec:simulations}

\subsection{Method}
\label{sec:method}

We use the {\tt Athena++} code \citep{Stone20} to compute the surface density and temperature profiles of disks with an embedded planet. Our simulations assume a 2D viscous circumstellar disk with constant $\alpha$ viscosity parameter \citep{Shakura73}. The temperature distribution is calculated by solving the time-dependent energy equation with the $\beta$ relaxation term \citep{Gammie01}. 
The planet is assumed to orbit circularly at a fixed orbital radius $r_{\rm p}$ and with the local Keplerian frequency at that radius, $\Omega_{\rm p}$.
We refer the reader to \citet{Ono25} for more details on our simulation method.

Our simulations evolve the 2D distributions of the disk surface density $\Sigma$, velocity ${\bm v}$, and specific internal energy $e$ (see equations (1), (2), and (5) of \citealt{Ono25}). The temperature $T$ is given by $T = e/c_{\rm v}$, where $c_{\rm v}$ is the specific heat at constant volume. Another useful measure of temperature is the isothermal sound speed $c_T$, given by 
\begin{equation}
c_{T}^2 = (\gamma -1)e = c_{\rm v}(\gamma-1)T,  
\label{eq:cT2}
\end{equation}
where $\gamma$ is the adiabatic index, which we take to be $\gamma = 1.4$ in our simulations. Since our simulations assume slow thermal relaxation (see below), the simulated disks behave nearly adiabatically, and therefore the actual sound speed is better approximated by the adiabatic one, $\sqrt{\gamma}c_T$, rather than by $c_T$ (see also section~\ref{sec:analytic} and appendix~\ref{sec:Kanagawa}).

With the $\alpha$ viscosity prescription, the disk's kinematic viscosity is given by $\nu = \alpha c_T^2/\Omega_{\rm K}$, where $\alpha$ is a dimensionless constant, which we call the viscosity parameter, and  $\Omega_{\rm K}(r)$ is the local Keplerian frequency.
To isolate the role of shock heating, we neglect viscous heating arising from Keplerian shear. This treatment is also motivated by the picture of MHD-driven accretion in cold protoplanetary disks, where accretional heating occurs predominantly near the disk surface and gives only a minor contribution to midplane heating \citep[e.g.,][]{Hirose11,Mori19}. Limitations of using the classical viscous model to protoplanetary disks are discussed in section~\ref{sec:alpha}.

The $\beta$ relaxation prescription forces the temperature at each position to relax toward its initial value $T_{\rm init}$ over an e-folding time of $\beta/(2\pi)$ local orbital periods, where $\beta$ is a dimensionless constant. This is implemented by adding the relaxation term   
\begin{equation}
Q_{\rm relax} = \frac{\Sigma c_{\rm v}\Omega_{\rm init}}{\beta}(-T+T_{\rm init})
\label{eq:Qrelax}
\end{equation}
to the energy equation (see equations (5) and (11) of \citealt{Ono25}), where
$\Omega_{\rm init}$ is the initial disk rotation frequency (see below). 
On the right-hand side of equation~\eqref{eq:Qrelax}, the first term accounts for cooling, while the second term represents heating from non-hydrodynamic sources, such as stellar irradiation. 
{Our simulations assume $\beta \gg 1$, representative of the inner $\lesssim$ 10 au of a dusty protoplanetary disk (see, e.g., figure~7 of \citealt{Ziampras20}; figure~11 of \citealt{Okuzumi22}). In this regime, radiative damping of planet-induced spiral waves, which becomes significant when $\beta \approx 0.1$--1 \citep[e.g.,][]{Miranda20a,Miranda20b,ZhangZhu20,Ziampras20b}, is negligible}.

The initial radial profiles of the surface density and temperature are given by $\Sigma(r) = \Sigma_{\rm init}(r) \equiv \Sigma_{\rm init,p}(r/r_{\rm p})^{-1}$ and $T(r) = T_{\rm init}(r) \equiv T_{\rm init,p}(r/r_{\rm p})^{-1/2}$, where $\Sigma_{\rm init,p}$ and $T_{\rm init,p}$ are constants.   
While $\Sigma_{\rm init,p}$ is an arbitrary constant in our simulations, $T_{\rm init,p}$ is chosen such that the initial isothermal sound speed at the planet's orbit, $c_{T, \rm init}(r_{\rm p})$,  satisfies
\begin{equation}
  c_{T, \rm init}(r_{\rm p}) = h_{\rm p}r_{\rm p}\Omega_{\rm p},    
\end{equation}
where $h_{\rm p}$ is a dimensionless parameter of the simulation and $r_{\rm p}\Omega_{\rm p}$ denotes the Keplerian speed at $r = r_{\rm p}$.
The length $h_{\rm p}r_{\rm p}$ represents the initial isothermal scale height at the planet's orbit, $c_T(r_{\rm p})/\Omega_{\rm p}$.
Throughout this paper, we refer to $h_{\rm p}$ simply as the initial disk aspect ratio at the planet's orbit.
The initial radial gas velocity $v_{r,\rm init}(r)$ is set to be the steady-state velocity in the viscous disk model with no planet, 
\begin{equation}
v_{r, \rm init} = -\frac{3\alpha c_{T}^2}{2r\Omega_{\rm K}}.
\label{eq:vrinit}
\end{equation}
The initial $\Sigma$ and $T$ profiles ensure a global viscous accretion flow in the absence of a planet, characterized by a radially constant mass accretion rate $\dot{M}_{\rm init} \equiv -2\pi r v_{r,\rm init}\Sigma_{\rm init} = 3\pi \alpha(r_{\rm p}h_{\rm p})^2 \Omega_{\rm p} \Sigma_{\rm init, p}$.
The initial disk rotation velocity $\Omega_{\rm init}(r)$ is determined from the radial force balance including the pressure gradient force. We  assume $h_{\rm p} \ll 1$ (see below), so that $\Omega_{\rm init} \approx \Omega_{\rm K}$.

The planet mass is gradually increased from zero to its final value of $M_{\rm p}$ over 50 orbits (see equation~(15) of \citealt{Ono25}). {The planetary gravitational force follows the standard Plummer-type prescription (see the ${\bm g}_{\rm p}$ term in equation~(6) of \citealt{Ono25}) with a smoothing length of 0.7 times the scale height \citep{Muller12}. Here, we adopt the adiabatic scale height, $\sqrt{\gamma}h_{\rm p}r_{\rm p}$, which is $\sqrt{\gamma}$ $(\approx 1.2)$ times larger than the isothermal scale height. Increasing the smoothing length by a factor of $2$ results in a planet-induced gap shallower by a comparable factor, but has virtually no effect on the specific entropy jumps across planet-induced spirals presented in the following section.}

\begin{table*}[t]
    \caption{List of simulation runs.}
    \label{tab:runs}
    \centering
    \begin{tabular}{lccccccc}
    \hline
    Run Name  & $M_{\rm p}/M_*$ & $h_{\rm p}$ & $M_{\rm p}/M_{\rm th}$ & $r_{\rm in}/r_{\rm p}$ & $r_{\rm out}/r_{\rm p}$ & $K$\\
    \hline 
    Run 1 & $10^{-4}$ & $0.05$ & $0.8$ & 0.73 & 1.4 & 14\\
    Run 2 & $10^{-4}$ & $0.03$ & $3.7$ & 0.6 & 1.6 & 180\\
    Run 3 & $10^{-3}$ & $0.07$ & $2.9$ & 0.54 & 1.8 & 260\\
    Run 4 & $10^{-3}$ & $0.05$ & $8.0$ & 0.51 & 2.2 & 1380\\    
    \hline
    \end{tabular}
\end{table*}
A useful unit for characterizing the planetary mass in the context of planet--disk interaction is the thermal mass $M_{\rm th}$, defined as \citep[e.g.,][]{GoodmanRafikov01}
\begin{equation}
M_{\rm th} \equiv h_{\rm p}^{3}M_*,
\label{eq:Mth}
\end{equation}
where $M_*$ is the stellar mass. The ratio $M_{\rm p}/(3M_{\rm th})$ represents the planet's Hill radius $(M_{\rm p}/3M_*)^{1/3}r_{\rm p}$ normalized by the disk's isothermal gas scale height at the planet's orbit, $r_{\rm p}h_{\rm p}$. When $M_{\rm p}/M_{\rm th} \gtrsim 3$, planet-induced spiral arms exhibit strong nonlinearity \citep[e.g.,][]{GoodmanRafikov01}.
{Following the convention in the literature, our definition of $M_{\rm th}$ uses the isothermal scale height; the thermal mass defined in terms of the adiabatic scale height differs from our $M_{\rm th}$ by a factor of $\gamma^{3/2} \approx 1.7$.}

The key dimensionless parameters of our simulations are the viscosity parameter $\alpha$, thermal relaxation parameter $\beta$, normalized planetary mass $M_{\rm p}/M_{\rm th}$, and initial aspect ratio at the planet's orbit, $h_{\rm p}$. In this study, we present four simulation runs, named Runs 1--4, with $\alpha = 10^{-3}$, $\beta = 10^2$, and different values of $M_{\rm p}$ and $h_{\rm p}$, as listed in table~\ref{tab:runs}. The values of the  normalized mass $M_{\rm p}/M_{\rm th}$ range between $0.8$ and 8. We fix the value of $\alpha$, as \citet{Ono25} have already explored how the entropy jumps across planet-induced spiral shocks scale with this parameter (see subsection~\ref{sec:simulationresults}). Since the entropy jumps are found to be insensitive to the $\beta$ parameter \citep{Ono25}, we also fix this parameter in this study. We note, however, that the gap temperature does depend on $\beta$, with a higher value of $\beta$ leading to a higher temperature, as the temperature is set by the balance between shock heating and thermal relaxation. Moreover, the use of a constant-$\beta$ relaxation prescription can destabilize the gap temperature structure when $\beta$ is large, as empirically found by \citet{Ono25}. In this study, we choose a $\beta$ value low enough to maintain temperature stability, which, however, results in only a moderate increase in the gap temperature. We discuss this issue further in subsection~\ref{sec:beta}. 

Our simulation domain covers $ r \in [r_{\rm bnd,in}, r_{\rm bnd,out}]$ in the radial direction and $\phi \in [-\pi, \pi]$ in the azimuthal direction, where $r$ and $\phi$ are the radial and azimuthal coordinates, respectively, {and the inner and outer radial boundaries are set at $r_{\rm bnd,in} = 0.4 r_{\rm p}$ and $r_{\rm bnd,out} = 9.88 r_{\rm p}$}. 
{All dynamical variables at the radial boundaries are fixed at their initial values. In addition, to reduce wave reflection at these boundaries, we follow \citet{DeValBorro06} and force the velocity near the boundaries to relax toward the initial value ${\bm v}_{\rm init}$ at an e-folding rate of 
\begin{equation}
    \eta = \min(x^2,1)
    \Omega_{\rm K}, 
\end{equation}
with
\begin{equation}
x = \begin{cases}
    \dfrac{1.25-r/r_{\rm bnd, in}}{1.25-1}, &1< {r}/{r_{\rm bnd, in}} < 1.25,\\[2mm]
    \dfrac{r/r_{\rm bnd, out}-0.84}{1-0.84},  &0.84< {r}/{r_{\rm bnd, out}} < 1.
\end{cases}
\end{equation}
Although our treatment of the radial boundaries follows a standard prescription, \citet{Dempsey20} point out that it has difficulty in capturing a global steady-state viscous accretion flow across the disk. This issue is discussed in subsection~\ref{sec:steady}.
A periodic boundary
condition is imposed on the azimuthal boundaries.}

Runs 1, 3, and 4 are evolved for 4000 planetary orbits, while Run 2 is extended to 6000 orbits to account for slower viscous diffusion (note that the viscosity $\nu \propto T$ initially scales as $h_{\rm p}^2$). In all runs, the surface density profile converges to a steady state within 10\% accuracy.

\subsection{Results}
\label{sec:simulationresults}

\begin{figure*}[t]
\begin{center}
\includegraphics[width=\hsize, bb=0 0 770 500]{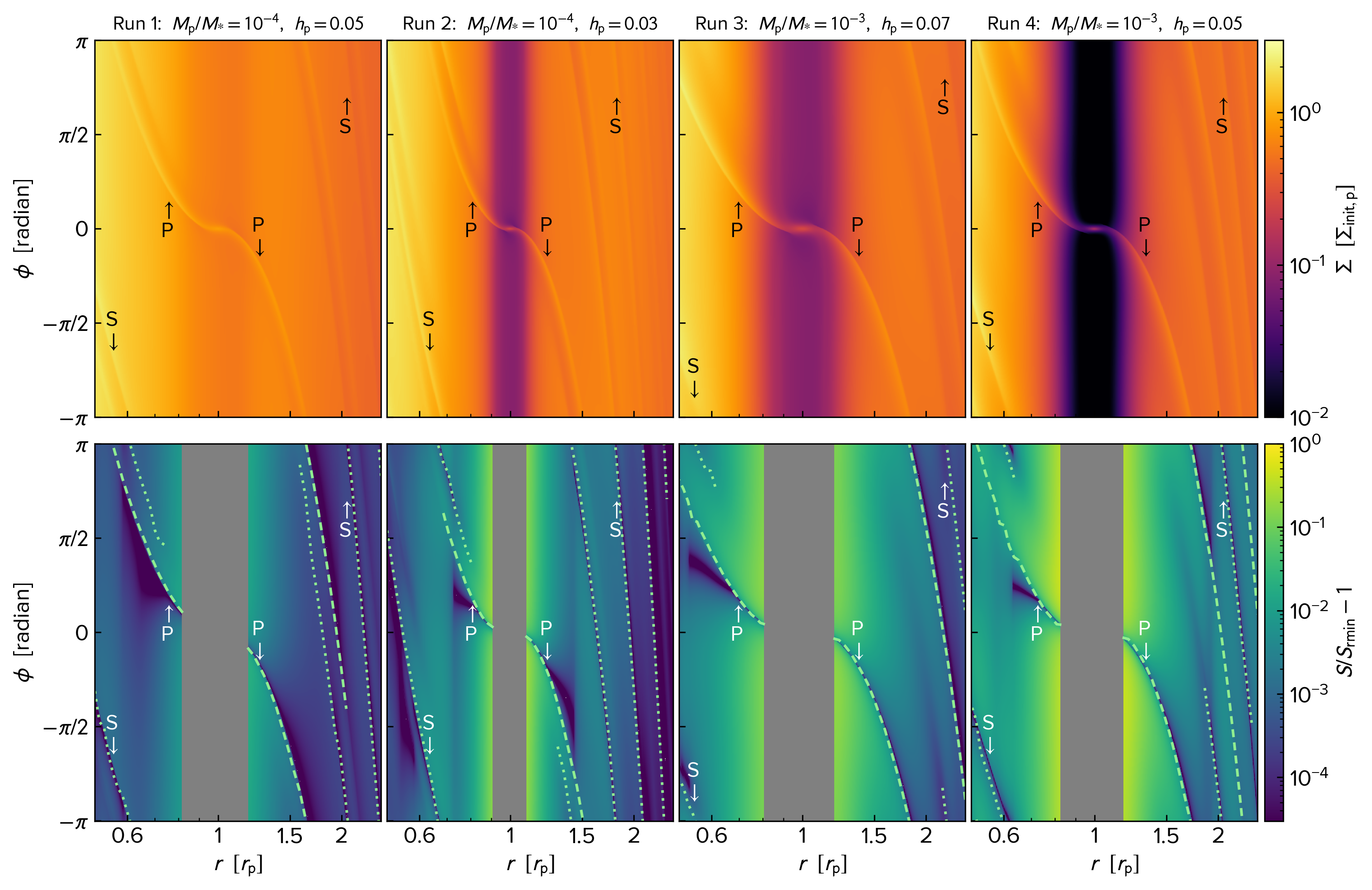}
\end{center}
\caption{{2D profiles of $\Sigma$ and ${\cal S}/{\cal S}_{\rm rmin} - 1$ (upper and lower panels, respectively)} from all simulation runs. {Here, ${\cal S}_{\rm rmin}(r)$ denotes the minimum value of the exponential entropy ${\cal S}$ at radial distance $r$.} The planet is located at $(r,\phi) = (r_{\rm p}, 0)$.
{The letters ``P'' and ``S'' label the primary and secondary spiral arms.} 
Dashed and dotted lines {in the lower panels} mark the locations of the primary and secondary shocks, respectively, {with $\delta > 5\times 10^{-5}$.} {Gray-shaded areas in the lower panels 
are excluded from streamline analysis.}
}
\label{fig:2D}
\end{figure*}

\begin{figure*}[t]
\begin{center}
\includegraphics[width=\hsize, bb=0 0 727.5 561.188125]{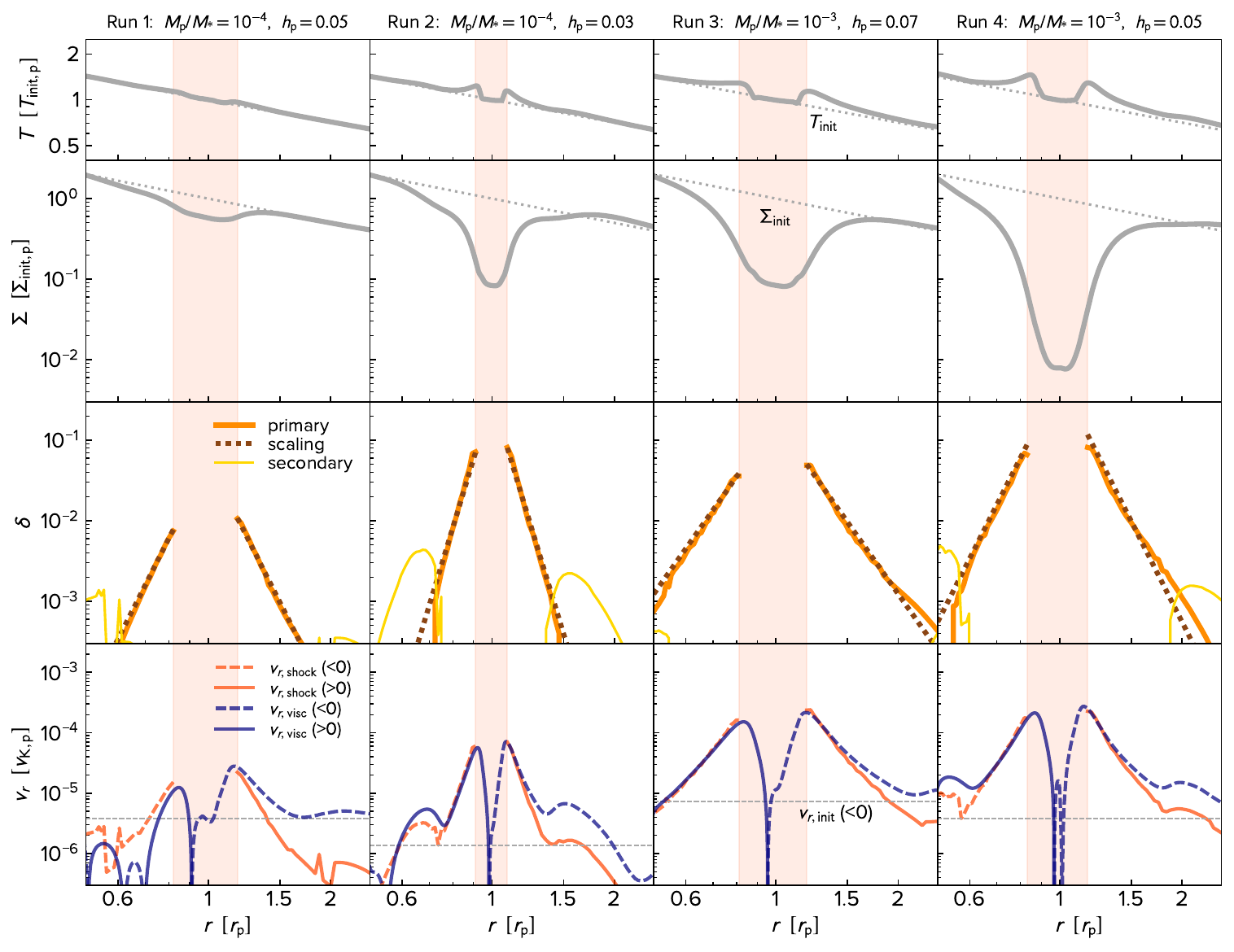}
\end{center}
\caption{Radial structure of the simulated disk models. Top two rows: azimuthally averaged temperature and surface density. Dotted lines show the initial profiles. Third row: dimensionless specific entropy jumps, $\delta$ (equation~\eqref{eq:delta}), across the primary and secondary shocks, obtained from streaming analysis. Bottom row: shock- and viscosity-induced radial gas velocity components, $v_{r,{\rm shock}}$ and $v_{r,{\rm visc}}$, estimated from equations~\eqref{eq:vr_shock} and \eqref{eq:vr_visc} (bottom row) for all simulation runs. Shaded regions around the planet's orbit are excluded from streamline analysis. Dotted lines in the first, second, and bottom rows represent the temperature, surface density, and radial velocity profiles before the planet's insertion. Dotted lines in the third row show the empirical scaling laws for the primary shocks' entropy jumps, equation~\eqref{eq:delta_formula} with $({\cal A}, {\cal B}) = (0.5, 0.5)$, as presented in subsection~\ref{sec:simulationresults}.
}
\label{fig:sigma_vr}
\end{figure*}

{The upper panels of figure~\ref{fig:2D} display} the 2D distribution of $\Sigma$ in the final (nearly steady-state) state from all simulation runs. 
The radial profiles of the azimuthally averaged temperature and surface density are shown in the upper two rows of figure~\ref{fig:sigma_vr}.
A gap in surface density is visible around the planetary orbit at $r = r_{\rm p}$, while temperature enhancements are evident at locations offset from the planet's orbit.

As discussed by a number of previous studies \citep[e.g.,][]{Kley99,Juhasz15,DongZhuRafikov15,ZhuDong15,FungDong15,Bae17,Bae18a,Bae18b}, a single planet can launch multiple spiral waves. This is confirmed in our simulations, which show two spiral arms both interior and exterior to the planetary orbit  (see {the upper panels of} figure~\ref{fig:2D}).  Following these previous studies, we call the arms directly attached to the planet the {\it primary} arms, and those that are not the {\it secondary} arms. The tertiary and quaternary inner arms, which can form at larger distances from the planet \citep[e.g.,][]{ZhuDong15,Bae18a,Bae18b}, are not visible because of the relatively small inner computational domain adopted in our simulations.
{The outer secondary arm is present in all our simulations (see also figure C.3 of \citealt{Ziampras20} for another example). However, this arm tends to vanish at higher values of $h_{\rm p}$  than those adopted in our simulations \citep{Bae18b,Miranda19a,Cimerman24}.}

Individual arms develop shocks as they propagate away from the planetary orbit. { To visualize the shocks, we present 2D maps of the exponential entropy, defined by ${\cal S} \equiv c_T^2/\Sigma^{\gamma+1}$, in the lower panels of figure~\ref{fig:2D}. In general, the specific entropy difference between arbitrary positions 1 and 2 is given by
$\Delta s_{2,1} = c_{\rm v}\ln({\cal S}_2/{\cal S}_1)$, where ${\cal S}_1$ and ${\cal S}_2$ are the values of ${\cal S}$ at those positions. A shock manifests as a discontinuity in ${\cal S}$.
A comparison between the upper and lower panels of figure~\ref{fig:2D} confirms that sharp entropy jumps indeed occur at the locations of the spiral density waves. 
We quantify the entropy jump across each shock using the following dimensionless quantity \citep[][see also equation~(26) of \citealt{Rafikov16}]{Ono25}, 
\begin{equation}
\delta 
\equiv \exp\left(\frac{\Delta s_{\rm shock}}{c_{\rm v}}\right)-1
= \frac{{\cal S}_{\rm pre}}{{\cal S}_{\rm post}}-1,
\label{eq:delta}
\end{equation}
where $\Delta s_{\rm shock} = c_{\rm v}\ln({\cal S}_{\rm post}/{\cal S}_{\rm pre})$ is the specific entropy jump across the shock, with ${\cal S}_{\rm pre}$ and ${\cal S}_{\rm post}$ being the pre-shock and post-shock values of ${\cal S}$, respectively. For $\delta \ll 1$, which holds in our simulations, $\delta$ approximates to $\Delta s_{\rm shock}/c_{\rm V}$. For this reason,} we call $\delta$ the dimensionless entropy jump.
In the following, we call the shocks associated with the primary and secondary arms the primary and secondary shocks, respectively. 

Following \citet{Ono25}, we detect these shocks by drawing streamlines of various initial orbital radii and search for increases in specific entropy along each line (for details, see subsection 2.3 of \citealt{Ono25}). 
Our streamline analysis excludes the region where spiral arms have not yet formed shocks. To determine where to exclude, we note that the temperature enhancements visible in the second row of Figure~\ref{fig:sigma_vr} result from heating by the primary shocks. We expect the primary waves to transition into shocks at locations where the temperature enhancements are peaked. Visual inspection shows that the peaks of the relative temperature enhancement $T/T_{\rm init}$ are located at $r - r_{\rm p} \approx \pm \varepsilon$, where $\varepsilon/r_{\rm p} \approx 0.18$, {0.09}, 0.19, and 0.17 for Runs 1, 2, 3, and 4, respectively {(see figure~\ref{fig:fheat} in appendix~\ref{sec:fheat})}. Based on this inspection, we exclude the region of $|r - r_{\rm p}| < \varepsilon$ from our streamline analysis. The estimated values of $\varepsilon$ are qualitatively consistent with the expectation that primary shocks form at larger distances when the planet is {less} massive or the sound speed (proportional to $h_{\rm p}$) is higher \citep{GoodmanRafikov01}.
The values of the ratio $\varepsilon/(h_{\rm p}r_{\rm p})$ indicate that the primary shocks develop at about $\pm 3$ isothermal scale heights away from the planet's orbit. The dashed and dotted lines {in the lower panels of} figure~\ref{fig:2D} mark the locations of the primary and secondary shocks, respectively, {detected by our streamline analysis.}

The third row of figure~\ref{fig:sigma_vr} shows $\delta$ across the primary and secondary shocks as functions of orbital radius. For all runs, the primary shocks dominate the total $\delta$ in the vicinity of the planet, while the secondary shocks dominate over the primary ones at sufficiently large distances from the planetary orbit. 
As $h_{\rm p}$ is increased, the secondary shocks develop farther from the planet because of the larger sound speed (see also \citealt{Ono25}).

The log--log plots in the third row of figure~\ref{fig:sigma_vr} 
suggest that the entropy jumps across the primary shocks follow a power-law dependence on radius $r$. 
The power-law nature of the primary shocks' entropy jumps was first noted by \citet{Ono25}, although their simulations were limited to $h_{\rm p} = 0.05$. 
\citet{Ono25} empirically found a common scaling 
$\delta \propto \alpha^{1/2}(M_{\rm p}/M_*)(r/r_{\rm p})^{f_1}$, where $f_1$ is a constant that is positive and negative for the inner ($r < r_{\rm p}$) and outer ($r > r_{\rm p}$) primary shocks, respectively. 
Our simulations with different values of $h_{\rm p}$ reveal that the slope $f_1$ actually depends on $h_{\rm p}$, with lower $h_{\rm p}$ leading to a steeper radial slope (see the third row of figure~\ref{fig:sigma_vr}).
In light of this observation, we propose a new {empirical} fit that explicitly incorporates the dependence of $\delta$ on $h_{\rm p}$,
\begin{equation}
\delta = {\cal A} \pfrac{\alpha}{h_{\rm p}}^{1/2}
\pfrac{M_{\rm p}}{M_{\rm th}}
\pfrac{r}{r_{\rm p}}^{-\xi {\cal B}/h_{\rm p}},
\label{eq:delta_formula}
\end{equation}
where $\xi(r) \equiv -1$ and 1 for $r < r_{\rm p}$ and $r > r_{\rm p}$, respectively, and ${\cal A}$ and ${\cal B}$ are numerical constants.
In this new fit, we normalize the planetary mass with the thermal mass $M_{\rm th}$ defined by equation~\eqref{eq:Mth}, rather than $M_*$, as $M_{\rm th}$ is a natural unit of the planetary mass for general values of $h_{\rm p}$ in the context of planet--disk interaction. Aside from this $M_{\rm p}/M_{\rm th}$ dependence, the new fit assumes $\delta \propto (\alpha/h_{\rm p})^{1/2}$,  instead of $\delta \propto \alpha^{1/2}$ adopted in the fit by \citet{Ono25}; as we show in subsection~\ref{sec:scaling}, this choice allows our gap model to reproduce a gap opening criterion found by previous studies. The assumed radial slope of $-{\cal B\xi}/h_p$ satisfies the observed properties that the slope is positive and negative for $r < r_{\rm p}$ and $r > r_{\rm p}$, respectively, and that the slope's magnitude increases with decreasing $h_{\rm p}$. 

We find that equation~\eqref{eq:delta_formula} with $({\cal A}, {\cal B}) = (0.5, 0.5)$, as shown by the dotted lines in the third row of figure~\ref{fig:sigma_vr}, provides a reasonable fit to $\delta$ at both the inner and outer primary shocks. Specifically, this choice reproduces the entropy jumps across the primary shocks over the range $0.7r_{\rm p} \lesssim r \lesssim 1.4r_{\rm p}$ with an error of less than $20\%$ for all four simulations.

\section{Relating spiral-induced disk heating to gap opening} \label{sec:bridge}
Shock heating raises the temperature at the gap edge above the initial temperature (the top row of figure~\ref{fig:sigma_vr}). 
The shocks also carve a surface density gap around the planet's orbit (figure~\ref{fig:2D} and the second row of figure~\ref{fig:sigma_vr}). In this section, we show that the radial gas flow driving gap opening can be described as a function of the entropy jump $\delta$ introduced in section~\ref{sec:simulations}.

Gap opening is due to the angular momentum deposition by the spiral shocks onto the background disk.
We begin by predicting the relation between the shock-induced radial gas velocity and $\delta$. 
Writing the rate of angular momentum deposition by spiral shocks per unit orbital radius as  $\Lambda_{\rm shock}(r)$, the induced radial mass flux is given by $F_{M,\rm shock} = 2\Lambda_{\rm shock}/(r\Omega_{\rm K})$ \citep[see, e.g., equation~(10) of][]{KanagawaTanaka15}, and the azimuthally averaged induced radial flow can be written as 
\begin{equation}
v_{r, \rm shock} \equiv \frac{F_{M,\rm shock}}{2\pi r\Sigma} =  \frac{\Lambda_{\rm shock}}{\pi r^2 \Omega_{\rm K}\Sigma}.
\label{eq:vr_shock_0}
\end{equation}
\citet{Rafikov16} and \citet{ArzamasskiyRafikov18} showed that $\Lambda_{\rm shock}(r)$ is related to the time-averaged shock heating rate per unit disk area, $Q_{\rm shock}$, through the following simple relation \citep[see also][]{GoodmanRafikov01}:
\begin{align}
\Lambda_{\rm shock}(r) 
&= \frac{2\pi r Q_{\rm shock}(r)}{\Omega_{\rm p}-\Omega_{\rm K}(r)}.
\label{eq:Lambda_shock}
\end{align}
The shock heating rate is in turn related to $\delta$ as \citep{Rafikov16,Ono25}
\begin{equation}
Q_{\rm shock}(r) = \frac{\Sigma c_{T}^2|\Omega_{\rm p}-\Omega_{\rm K}(r)|}{2\pi(\gamma-1)}\delta.
\label{eq:Qshock}
\end{equation}
We thus have
\begin{equation}
\Lambda_{\rm shock}
= \xi\frac{r\Sigma c_{T}^2}{\gamma-1}\delta
\label{eq:Lambda_shock_2}
\end{equation}
and
\begin{equation}
{v}_{r,{\rm shock}} = \xi\frac{c_{T}^2 }{\pi(\gamma-1) r\Omega_{\rm K}}\delta,
\label{eq:vr_shock}
\end{equation}
where we have used that ${\rm sgn}(\Omega_{\rm p}-\Omega_{\rm K}(r)) = \xi(r)$.
Equation~\eqref{eq:vr_shock} indicates that ${v}_{r,{\rm shock}}$ is directed away from the planet's orbit, consistent with the picture of gap opening. 

To demonstrate that $v_{r,\rm shock}$ given by equation~\eqref{eq:vr_shock} indeed describes the gas flow driving gap opening, we consider steady-state gaps, where the shock-induced flow should counterbalance the viscosity-induced flow that replenishes the gas into the gap \citep{Lubow06}. 
The viscosity-induced radial velocity is given by \citep{Lynden-Bell74}
\begin{equation}
{v}_{r,{\rm visc}} = -\frac{3\alpha c_{T}^2}{r\Omega_{\rm K}}\frac{\partial\ln(r^2\alpha c_{T}^2 \Sigma)}{\partial \ln r}. 
\label{eq:vr_visc}
\end{equation}
The bottom row of figure~\ref{fig:sigma_vr} compares the radial profiles of ${v}_{r,{\rm shock}}$ and ${v}_{r,{\rm visc}}$ estimated from equations~\eqref{eq:vr_shock} and \eqref{eq:vr_visc}, respectively. Here, $\delta$ in ${v}_{r,{\rm shock}}$ is taken to  be the sum of the entropy jumps across the primary and secondary shocks measured by the streamline analysis, whereas $\Sigma$ and $c_T$ in ${v}_{r,{\rm shock}}$ and ${v}_{r,{\rm visc}}$ are taken from the azimuthally averaged surface density and temperature obtained from the simulations\footnote{We note that neither ${v}_{r,{\rm shock}}$ nor ${v}_{r,{\rm visc}}$ is a directly measurable radial gas velocity in the gap region. Rather, they should be interpreted as contributions to the azimuthally averaged radial velocity arising from distinct physical processes.}. As expected, we find that  $v_{r,\rm shock}$ given by equation~\eqref{eq:vr_shock}  counterbalances the viscosity-induced velocity in the gap region, where the surface density is appreciably reduced from its initial value.

\section{Predicting the gap temperature and surface density structure from the entropy jump profile} 
\label{sec:semi}

In the previous section, we used the surface density profiles obtained from simulations to confirm that the balance between $v_{r,\rm shock}$ and $v_{r,\rm visc}$ holds in the gap region. Here, we take the opposite approach and {\it predict} the surface density profile in the gap by assuming the balance $v_{r,\rm shock} + v_{r,\rm visc} = 0$. A similar analysis was performed by \cite{Lubow06}, but they assumed instantaneous damping of the planet-induced spirals. In contrast, equation~\eqref{eq:vr_shock} accounts for the fact that angular momentum deposition by spirals occurs only after they shock, which is essential for accurately modeling the gas structure \citep{KanagawaTanaka15}. Furthermore, we aim to simultaneously predict the disk temperature profile by taking shock heating into account. {In this and the following sections, we demonstrate that this approach yields self-consistent analytic models for the temperature and surface density profiles around the gap.}

We assume that the balance $v_{r,\rm shock} + v_{r,\rm visc} = 0$ approximately holds in a region of $r_{\rm in} < r < r_{\rm out}$ (excluding the planet's very vicinity of $|r-r_{\rm p}| < \varepsilon$), where $r_{\rm in}$ ($< r_{\rm p}$) and $r_{\rm out}$ ($> r_{\rm p}$) represent the inner and outer edges of that region, respectively. The balance $v_{r,\rm shock} + v_{r,\rm visc} = 0$ can be rewritten as
\begin{equation}
\frac{\partial\ln(r^2\alpha c_T^2 \Sigma)}{\partial \ln r} = \xi\frac{\delta}{3\pi(\gamma-1)\alpha}.    
\label{eq:balance}
\end{equation}
Integrating this from $r_{\rm in}$ to $r (<r_{\rm p}- \varepsilon)$ yields
\begin{equation}
 \Sigma(r) = 
 \dfrac{(r^2\alpha c_T^2 \Sigma)_{r_{\rm in}}}{r^2\alpha c_T^2 }
 \exp\left( -\dint_{r_{\rm in}}^r \dfrac{\delta(r')dr' }{3\pi(\gamma-1)\alpha r'} \right)
 \label{eq:Sigma_semi_in}
\end{equation}
for $r_{\rm in} < r < r_{\rm p} - \varepsilon$. Similarly, we obtain 
\begin{equation}
 \Sigma(r) = 
 \dfrac{(r^2\alpha c_T^2 \Sigma)_{r_{\rm out}}}{r^2\alpha c_T^2 }\exp\left(-\dint^{r_{\rm out}}_r \dfrac{\delta(r')dr' }{3\pi(\gamma-1)\alpha r'} \right)
 \label{eq:Sigma_semi_out}
\end{equation}
for $ r_{\rm p} + \varepsilon < r < r_{\rm out}$. 
Importantly, if $\delta$ and $T$ are independent of $\Sigma$, equations~\eqref{eq:Sigma_semi_in} and \eqref{eq:Sigma_semi_out} explicitly determine the surface density profile given the boundary positions, $r_{\rm in}$ and $r_{\rm out}$, and the boundary values, $\Sigma(r_{\rm in})$ and $\Sigma(r_{\rm out})$.

{To determine $r_{\rm in}$ and $r_{\rm out}$, we note that the surface density profiles observed in our simulations approach $\Sigma \approx \Sigma_{\rm init}$ at large distances from the planet (as discussed in subsection~\ref{sec:steady}, this  result could be affected by our choice of the radial boundary conditions and by a finite computational time).} In these regions, the radial gas velocity can be approximated by the steady-state velocity prior to the planet's insertion, $v_{r, \rm init}$ (equation~\eqref{eq:vrinit}), and the shock-induced radial velocity is much smaller than this velocity. Therefore, the gap edge locations, $r_{\rm in}$ and $r_{\rm out}$, can be defined as where $|v_{r,\rm shock}|=|v_{r, \rm init}|$. 
Combining equations~\eqref{eq:vrinit} and \eqref{eq:vr_shock}, this condition becomes
\begin{equation}
\delta(r_{\rm in,out}) =  \frac{3\pi\alpha(\gamma-1)}{2}
\approx 1.9 \alpha.
\label{eq:delta_inout}
\end{equation} 
Table~\ref{tab:runs} lists the values of $r_{\rm in}$ and $r_{\rm out}$ for all runs, derived by evaluating $\delta$ as the sum of the entropy jumps across the primary and secondary shocks obtained from the streamline analysis.
Our gap model simply sets $\Sigma = \Sigma_{\rm init}$ at $r \leq r_{\rm in}$ and $r \geq r_{\rm out}$.

Since $c_T^2 \propto T$, equations~\eqref{eq:Sigma_semi_in} and \eqref{eq:Sigma_semi_out} require knowledge of the radial temperature profile. Here, we predict the steady-state, azimuthally averaged temperature, using the analytic expressions for the shock heating rate $Q_{\rm shock}$ (equation~\eqref{eq:Qshock}) and thermal relaxation rate $Q_{\rm relax}$ (equation~\eqref{eq:Qrelax}). If local heating and relaxation exactly balanced  (i.e., $Q_{\rm shock} + Q_{\rm relax} = 0$), the temperature would be given by \citep{Ono25}
\begin{equation}
 T = T_{\rm init} \left( 1- \frac{\beta |1-\Omega_{\rm p}/\Omega_{\rm K}|}{2\pi}f_{\rm heat}\delta\right)^{-1},
 \label{eq:T_predict}
\end{equation}
with $f_{\rm heat} = 1$. However, we find that this simple model slightly overpredicts the temperature enhancements observed in our simulations ({see appendix~\ref{sec:fheat}}). This could be because this model neglects adiabatic cooling and advective heat transport around shocks. To account for these additional cooling processes, we have introduced an ad hoc dimensionless parameter $f_{\rm heat} < 1$, which effectively reduces the shock heating rate. {As shown in appendix~\ref{sec:fheat}, choosing $f_{\rm heat} = 0.8$ best reproduces the temperature enhancements in all our simulation runs.}

\begin{figure*}[t]
\begin{center}
\includegraphics[width=\hsize, bb=0 0 727.5 283.988125]{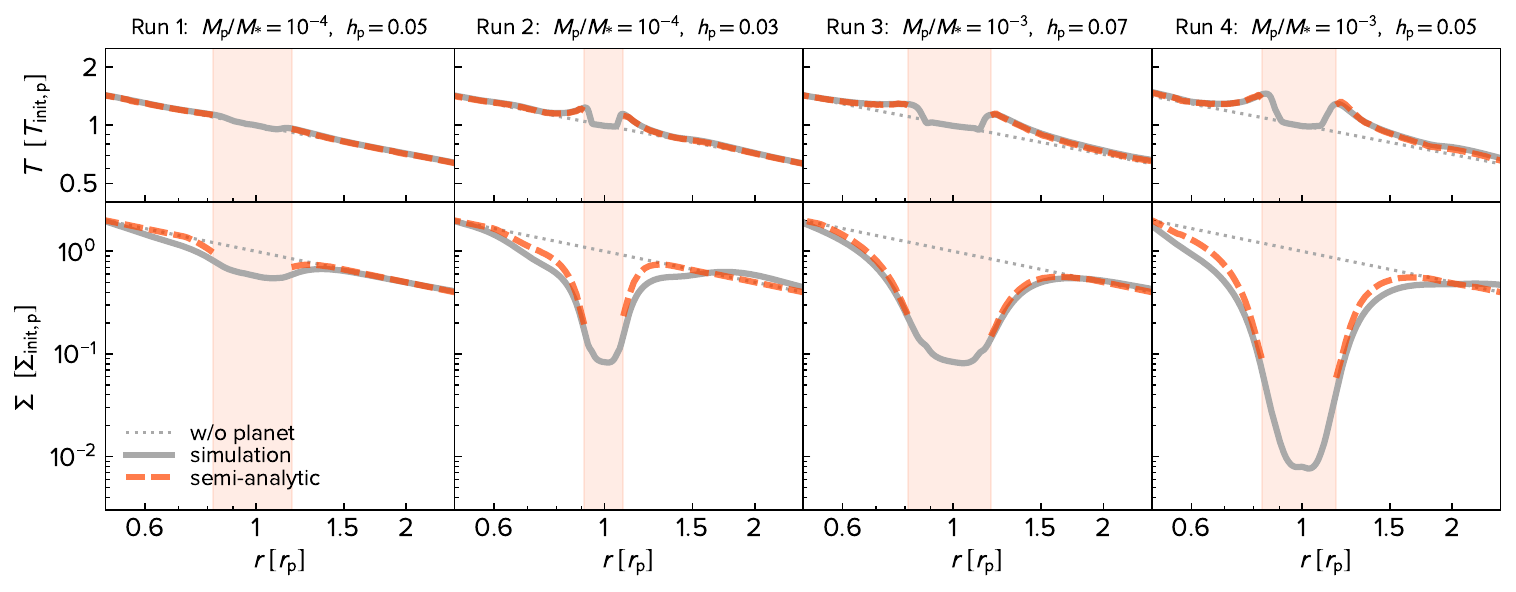}
\end{center}
 \caption{Comparison between simulations and the semi-analytic model (section~\ref{sec:semi}). The upper and lower panels show the radial temperature and surface density profiles, respectively. Solid lines represent azimuthally averaged profiles from four 2D simulations, Dashed lines show semi-analytic predictions based on equations~\eqref{eq:Sigma_semi_in}--\eqref{eq:T_predict}, using numerical $\delta$ profiles for primary and secondary spiral shocks obtained directly from streamline analysis.
 Shaded regions around the planet's orbit are excluded from streamline analysis. Dotted lines represent the profiles before the planet's insertion.
 }
\label{fig:semi}
\end{figure*}
The set of equations~\eqref{eq:Sigma_semi_in}--\eqref{eq:T_predict} predicts $\Sigma$ and $T$ once the radial profile of $\delta$ is specified. Figure~\ref{fig:semi} shows the predictions for our simulation runs using the sum of the entropy jumps across primary and secondary shocks obtained from the streamline analysis. We refer to these predictions as semi-analytic because rely on simulation data for $\delta$. These semi-analytic predictions are available at $|r - r_{\rm p}| >\varepsilon$, where we apply streamline analysis.
We find that the semi-analytic predictions match the azimuthally averaged surface density profiles from the simulations to within an accuracy of approximately 40\%. {With} $f_{\rm heat} = 0.8$, the temperature profiles are also reproduced with an accuracy {of $\lesssim$ 4\%  (see also appendix~\ref{sec:fheat})}.

{
We note that the model presented here neglects the net radial mass flux $F_{\rm M}$ across the gap by assuming the near balance $v_{r,\rm shock} + v_{r,\rm visc} \approx 0$. If we account for net accretion, we obtain
\begin{equation}
\frac{\partial\ln(r^2\alpha c_T^2 \Sigma)}{\partial \ln r} = \xi\frac{\delta}{3\pi(\gamma-1)\alpha} + \frac{\Omega_{\rm K}\dot{M}}{6\pi\alpha c_T^2 \Sigma}    
\label{eq:balance2}
\end{equation}
instead of equation~\eqref{eq:balance}, where $\dot{M} \equiv -F_{\rm M}$ is the mass accretion rate.
If we additionally assume that $\dot{M}$ is radially constant, equation~\eqref{eq:balance2} specifies a steady-state distribution of $c_T^2\Sigma$ with net accretion. 
Unlike equation~\eqref{eq:balance}, equation~\eqref{eq:balance2} has the advantage of being applicable across the gap edges.  
However, we do not use equation~\eqref{eq:balance2} in this study because {our simulations do not achieve a radially constant $\dot{M}$  (see section~\ref{sec:steady}).  Moreover, when $\delta$ is a power-law function of $r$, equations~\eqref{eq:Sigma_semi_in} and \eqref{eq:Sigma_semi_out} reduce to simple closed-form expressions (see equations~\eqref{eq:Sigma_analytic_in} and \eqref{eq:Sigma_analytic_out} in section~\ref{sec:analytic}), whereas the solution for equation~\eqref{eq:balance2} involves incomplete Gamma functions}.
}

\section{A consistent analytic model for the gap temperature and surface density profiles }\label{sec:analytic}
The semi-analytic predictions presented in section~\ref{sec:semi} still rely on the entropy jump profiles directly obtained from the simulations.
In this section, we present a fully analytic model for the gap profiles, utilizing the empirical scaling for $\delta$, equation~\eqref{eq:delta_formula}, presented in section~\ref{sec:simulationresults}.
To maintain consistency with the simulations from which this scaling was derived, the following calculations assume a radially constant $\alpha$.

Equation~\eqref{eq:delta_formula} provides a good estimate for the entropy jumps across the primary shocks, which are primarily responsible for shaping the gap structure around the planetary orbit. Furthermore, equation~\eqref{eq:delta_formula} is expressed as a power law of $r$ and does not involve $\Sigma$ or $T$, making the radial integration in equations~\eqref{eq:Sigma_semi_in} and \eqref{eq:Sigma_semi_out} analytically tractable. Specifically, 
substituting equation~\eqref{eq:delta_formula} into equations~\eqref{eq:Sigma_semi_in} and \eqref{eq:Sigma_semi_out}
yields 
\begin{equation}
 \Sigma (r) = \frac{(r^2 c_T^2 \Sigma)_{r_{\rm in}}}{r^2 c_T^2 } 
 \exp\left( \dfrac{h_{\rm p}[\delta(r_{\rm in})-\delta(r)]}{3\pi{\cal B}(\gamma-1)\alpha }  \right)
\label{eq:Sigma_analytic_in}
\end{equation}
for $r_{\rm in} < r < r_{\rm p}$, and 
\begin{equation}
 \Sigma (r) = \frac{(r^2c_T^2 \Sigma)_{r_{\rm out}}}{r^2 c_T^2 }  
 \exp\left( \dfrac{h_{\rm p}[\delta(r_{\rm out})-\delta(r)]}{3\pi{\cal B}(\gamma-1)\alpha } \right)
\label{eq:Sigma_analytic_out} 
 \end{equation}
for $r_{\rm p} < r < r_{\rm out}$. 
If $T$ is independent of $\Sigma$, equations~\eqref{eq:Sigma_analytic_in} and \eqref{eq:Sigma_analytic_out} provide closed-form expressions for the radial surface density distribution in these regions. 

As noted in subsection~\ref{sec:simulationresults}, equation~\eqref{eq:delta_formula} does not apply to the gap center where shocked spirals are absent.  Therefore, extrapolating equations~\eqref{eq:Sigma_analytic_in} and \eqref{eq:Sigma_analytic_out}  toward $r \to r_{\rm p}$ leads to an overestimation of the gap depth. Here, we address this issue by simply assuming a floor surface density, $\Sigma_{\rm floor}$, at the gap and setting $\Sigma(r)  =  \Sigma_{\rm floor}$ wherever $\Sigma(r)$ derived from equation~\eqref{eq:Sigma_analytic_in} or \eqref{eq:Sigma_analytic_out} falls below this floor value. For the floor value, we adopt the estimate proposed by \citet[][see also~\citealt{Duffell13}]{KanagawaMuto15,KanagawaTanaka15}, but with a modification (see appendix \ref{sec:Kanagawa}),
\begin{equation}
\Sigma_{\rm floor} = \frac{\Sigma_{\rm init}}{1+0.04K}, 
\label{eq:Sigma_floor}
\end{equation}
\begin{equation}
\quad K =\frac{1}{\alpha}\pfrac{M_{\rm p}}{M_*}^2 (\sqrt{\gamma}h_{\rm p})^{-5}.    
\label{eq:Kanagawa}
\end{equation}
Here, $\sqrt{\gamma}h_{\rm p}$ stands for the ratio between the adiabatic sound speed $\sqrt{\gamma}c_{T}$ (see subsection~\ref{sec:method}) and Keplerian speed at $r=r_{\rm p}$. The factor $\sqrt{\gamma}$ is absent in the original expression by \citet{KanagawaTanaka15} as they assumed locally isothermal disks. In appendix~\ref{sec:Kanagawa}, we show that replacing $h_{\rm p}$ with $\sqrt{\gamma}h_{\rm p}$ improves the accuracy of the floor value estimates for our nearly adiabatic ($\beta =100\gg 1$) simulations.
{The values of $K$ for Runs 1--4 are given in table~\ref{tab:runs}.}
As in the semi-analytic model, we compute the gap edge locations $r_{\rm in}$ and $r_{\rm out}$ from equation~\eqref{eq:delta_inout}, but here using equation~\eqref{eq:delta_formula}. Combining the two equations yields  explicit expressions for $r_{\rm in}$ and $r_{\rm out}$, 
\begin{equation}
\frac{r_{\rm in}}{r_{\rm p}} = \left(\frac{2{\cal A}}{3\pi(\gamma-1)(\alpha h_{\rm p})^{1/2}}\frac{M_{\rm p}}{M_{\rm th}}\right)^{-h_{\rm p}/{\cal B}},
\label{eq:rin}
\end{equation}
\begin{equation}
\frac{r_{\rm out}}{r_{\rm p}} 
= \pfrac{r_{\rm in}}{r_{\rm p}}^{-1}.
\label{eq:rout}
\end{equation}
Outside the gap (i.e., $r < r_{\rm in}$ or $r > r_{\rm out}$), we set $\Sigma = \Sigma_{\rm init}$. 
A consistent temperature profile is derived from equation~\eqref{eq:T_predict} by setting $\delta = 0$ at the gap center (where $\Sigma = \Sigma_{\rm floor}$) and applying equation~\eqref{eq:delta_formula} elsewhere.

\begin{figure*}[t]
\begin{center}
\includegraphics[width=\hsize, bb=0 0 727.5 283.988125]{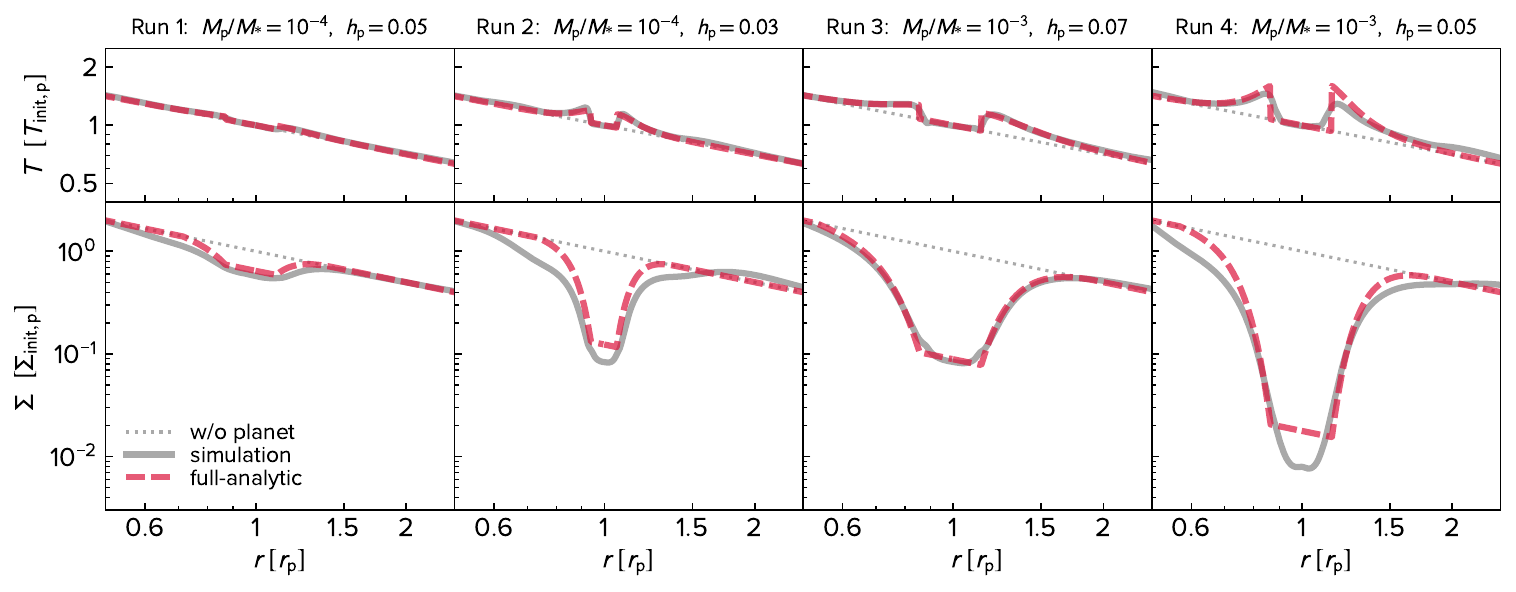}
\end{center}
\caption{Comparison between simulations and the full-analytic model (section~\ref{sec:analytic}). The upper and lower panels show the radial temperature and surface density profiles, respectively. Solid lines represent azimuthally averaged profiles from four 2D simulations. Dashed lines show full-analytic predictions ($T$ from equations~\eqref{eq:delta_formula} and \eqref{eq:T_predict}, and $\Sigma$ from equations~\eqref{eq:Sigma_analytic_in}--\eqref{eq:Kanagawa}) based on the empirical scaling law for the primary shocks' $\delta$ profiles, equation~\eqref{eq:delta_formula} with $({\cal A}, {\cal B}) = (0.5, 0.5)$. Dotted lines represent the profiles before the planet's insertion, $T = T_{\rm init}$ and $\Sigma = \Sigma_{\rm init}$.
}
\label{fig:analytic}
\end{figure*}
Figure~\ref{fig:analytic} compares the full-analytic predictions for the azimuthally averaged temperature and surface density profiles with those directly obtained from our simulations. For Runs 1 and 3, the full-analytic predictions are as accurate as the semi-analytic predictions (see figure~\ref{fig:semi}), both in terms of the temperature and surface density profiles. For Runs 2 and 4, the full-analytic model overepredicts the surface density at the outskirts of the gap by up to approximately 70\%. This is because the secondary shocks, which are neglected in the full-analytic model, provide $\delta$ values high enough to carve a gap ($\delta \gtrsim 3\pi\alpha(\gamma-1)/2 \sim 10^{-3}$) in these runs (see the third row of figure~\ref{fig:sigma_vr}). The floor value $\Sigma_{\rm floor}$ set by equation~\eqref{eq:Kanagawa} reproduces the surface density at the gap center in Runs 1--3 to within 20\% accuracy; for Run 4,  equation~\eqref{eq:Kanagawa} overpredicts the floor surface density by a factor of 2. The temperature distributions are reproduced with a relative error of less than 20\%.

The gap model constructed here is primarily intended to predict self-consistent temperature and surface density structures, rather than to provide a highly accurate surface density profile. Nevertheless, as shown in appendix~\ref{sec:Kanagawa}, we find that the surface density profiles predicted by our model are as accurate as, or even more accurate than, those of the widely used analytic gap model by \citet{KanagawaTanaka17}.

\section{Discussion}
\label{sec:discussion}
In this section, we discuss the validity and limitations of the assumptions adopted in this work and present prospects for future work. 

\subsection{Toward a deeper understanding of the  entropy jump scaling}
\label{sec:scaling}

The analytic model presented in section~\ref{sec:analytic} largely relies on the empirical scaling law for the entropy jumps across the primary shocks, given by equation~\eqref{eq:delta_formula}, which was derived heuristically in subsection~\ref{sec:simulationresults} as an extension of the expression proposed by \citet{Ono25}. While this formula provides a good fit to both our simulation results and those of \citet{Ono25}, it has yet to be derived from first principles governing the nonlinear propagation of spiral density waves, limiting our understanding of the physical origins of its parameter dependencies.
For instance, \citet{Ono25} empirically found that $\delta$ scales with viscosity as $\alpha^{1/2}$, but a physical explanation for this dependence remains lacking.

Although deriving the scaling law for $\delta$ from  first principles is beyond the scope of the present work, we can demonstrate that the assumed scaling reproduces an important property of planet-induced gaps already derived in the literature. \citet{Duffell13}, \citet{FungShi14}, and \citet{KanagawaMuto15,KanagawaTanaka15} have shown both analytically and numerically that the gap depth, $\Sigma_{\rm floor}/\Sigma_{\rm init}$, depends on the $K$ parameter defined by equation~\eqref{eq:Kanagawa}. Specifically, equation~\eqref{eq:Sigma_floor} indicates that a planet carves a deep gap of $\Sigma_{\rm floor}/\Sigma_{\rm init} \ll 1$ when $K \gtrsim 100$. In what follows, we show that our analytic gap model, based on equation~\eqref{eq:delta_formula}, recovers this deep-gap criterion.

In our analytic gap model, equations~\eqref{eq:Sigma_analytic_in} and \eqref{eq:Sigma_analytic_out} describe how the surface density decreases from the gap's edge to its center.
As $\delta$ is peaked at $r \approx r_{\rm p} \pm \varepsilon$, the exponential factors in equations~\eqref{eq:Sigma_analytic_in} and \eqref{eq:Sigma_analytic_out} indicate that a significant reduction in $\Sigma$ near the planet's orbit occurs when 
\begin{equation}
\frac{h_{\rm p}\delta(r_{\rm p} \pm \epsilon)}{3\pi {\cal B}(\gamma-1)\alpha} \gtrsim 1.
\label{eq:opening_1}
\end{equation}
For the value of $\delta(r_{\rm p} \pm \epsilon)$, equation~\eqref{eq:delta_formula} predicts
\begin{equation}
 \delta(r_{\rm p} \pm \epsilon) =  {\cal A} \pfrac{\alpha}{h_{\rm p}}^{1/2}
\pfrac{M_{\rm p}}{M_{\rm th}}
\left(1 \pm x h_{\rm p}\right)^{\mp{\cal B}/h_{\rm p}},
\end{equation}
where $x \equiv \varepsilon/(r_{\rm p}h_{\rm p})$; our simulations show that $x \sim 3$ (see subsection~\ref{sec:simulationresults}). 
Since ${\cal B} \sim x \sim O(1)$ and $h_{\rm p} \ll 1$, 
we can eliminate the $h_{\rm p}$ dependence of $\left(1 \pm x h_{\rm p}\right)^{\mp{\cal B}/h_{\rm p}}$ by approximating this factor with $e^{-{\cal B}x}$.
We then have
\begin{equation}
 \delta(r_{\rm p} \pm \epsilon) \approx 0.1 \pfrac{\alpha}{h_{\rm p}}^{1/2}
\pfrac{M_{\rm p}}{M_{\rm th}}
\end{equation}
for ${\cal A} = {\cal B} = 0.5$ and $x \sim 3$. This estimate reproduces our simulation results (the third row of figure~\ref{fig:sigma_vr}) to within 40\% accuracy. 
Using this and equation~\eqref{eq:Mth}, the deep-gap criterion, equation~\eqref{eq:opening_1},
can be rewritten as 
\begin{equation}
\frac{M_{\rm p}}{\alpha^{1/2} h_{\rm p}^{5/2}M_*} \gtrsim \frac{3\pi {\cal B}(\gamma-1)}{0.1} \approx 20.
\label{eq:opening_2}
\end{equation}
The left-hand side of the above inequality approximates to $\gamma^{5/4}K^{1/2}$, where $K$ is given by equation~\eqref{eq:Kanagawa}. Thus, equation~\eqref{eq:opening_2} leads to the final expression for our deep-gap criterion, $K \gtrsim 200$. This agrees with the criterion predicted from equation~\eqref{eq:Sigma_floor}. 

The above exercise largely justifies our choice of the parameter dependencies and radial scaling of $\delta$ adopted in equation~\eqref{eq:delta_formula}.
Equation~\eqref{eq:opening_2} would fail to recover the deep-gap criterion in terms of $K$ alone if, for instance, the $(\alpha/h_{\rm p})^{1/2}$ prefactor were missing from equation~\eqref{eq:delta_formula}.
In equation~\eqref{eq:opening_2}, the $\alpha^{-1/2}$ factor arises from both the $\alpha$ dependence of the viscosity-induced radial velocity, $v_{r,\rm visc}$ (equation~\eqref{eq:vr_visc}), and the $\alpha^{1/2}$ dependence of $\delta$. 
Physically, the $K$ parameter originates from the balance between
the Lindblad excitation torque and viscous angular momentum flux in the gap region \citep{Duffell13,FungShi14,KanagawaMuto15,KanagawaTanaka15}. This suggests that the $\alpha^{1/2}$ prefactor in $\delta$ may reflect a reduction in the Lindblad excitation torque due to the decreased surface density deep within the gap, where viscosity affects the surface density profile (S. Sato \& H. Tanaka 2025, private communication). This interpretation remains speculative and should be tested in future studies.

\subsection{Do our models capture steady states?}
\label{sec:steady}

{
In this study, we ran simulations until the surface density profile within and around the gap reached a nearly steady state. The gap structure in this state is characterized by a near balance between shock- and viscosity-induced radial flows (see sections~\ref{sec:bridge} and \ref{sec:semi}). However, this condition does not imply that a global steady-state  accretion flow, defined by a {\it net} mass flux $F_{\rm M} \equiv 2\pi r v_r \Sigma$ that is radially uniform and constant in time, has been established. To illustrate this, the top and middle panels of figure~\ref{fig:time} show the azimuthally averaged surface density and mass flux from Run 4 at different times. As explained in section~\ref{sec:method}, the initial surface density profile follows a global steady-state solution in the absence of a planet, with $F_{\rm M} = - \dot{M}_{\rm init}$ constant in $r$. In this run, the gas surface density relaxes to a nearly steady state within 4000 planetary orbits after the planet's insertion. However, the net mass flux does not return to a radially constant profile within this timescale. 
\begin{figure}[t]
\begin{center}
\includegraphics[width=\hsize, bb= 0 0 340.6734375 505.748125 ]{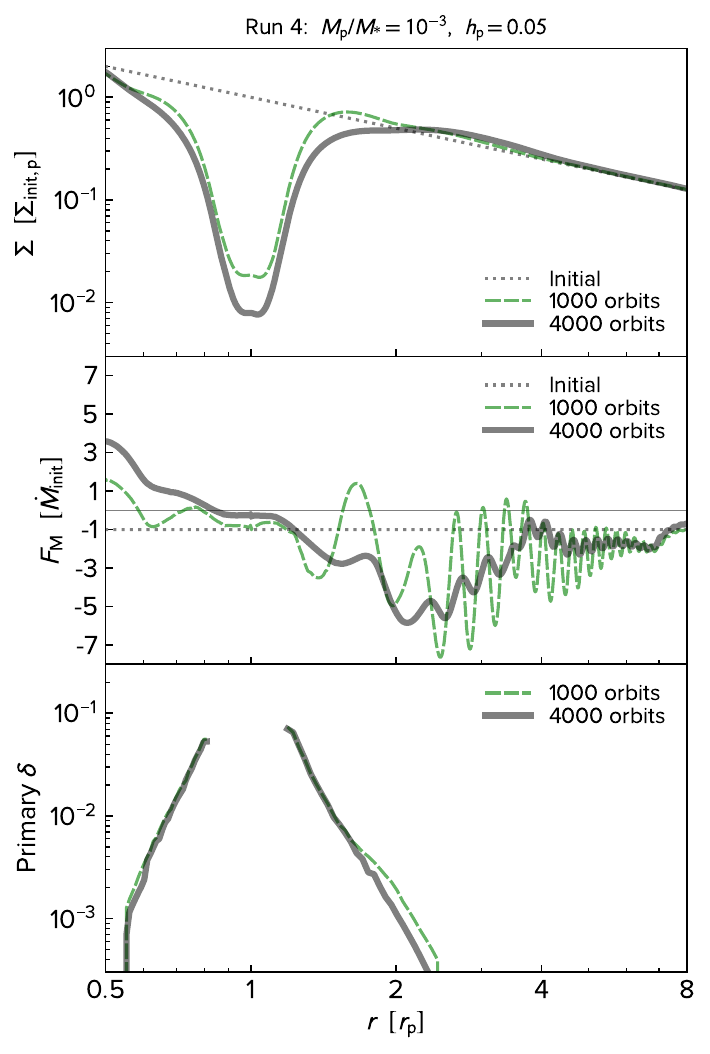}
\end{center}
\caption{Azimuthally averaged surface density and net radial mass flux (top and middle panels, respectively), and the sum of the entropy jumps at the primary and secondary shocks (bottom panel) in Run 4 at different times. Inward accretion corresponds to $F_{\rm M} < 0$.
}
\label{fig:time}
\end{figure}

Nevertheless, the gap models presented in sections~\ref{sec:semi} and \ref{sec:analytic} successfully reproduce the surface density profiles around a planet. This is because our models assume a near balance between the shock- and viscous-induced radial mass fluxes within the gap, such that each is significantly larger in magnitude than the net flux there. This assumption is approximately satisfied in the gaps formed in our simulations (see figure~\ref{fig:sigma_vr}). 

Still, the modeled gap profile depend on assumptions about the global accretion structure outside the gap region, through the junction conditions at the gap boundaries, $r = r_{\rm in}$ and $r = r_{\rm out}$. In our models, the gap boundaries have been chosen to be the locations where the shock-induced mass flux falls below the steady-state accretion flux expected in the absence of the planet. This choice is reasonable for our simulations, where $\Sigma$ approaches $\Sigma_{\rm init}$ far from the planet. However, \citet{Dempsey20} argue that the true surface density profile in a global viscous steady state should exhibit a pileup exterior to the planetary orbit. This is because the net torque exerted by the planet on the disk modifies the angular momentum flux across its orbit. \citet{Dempsey20} show that in the global viscous steady state, where $F_M$ is radially constant, the disk surface density at $r \gg r_{\rm p}$ becomes higher than that at $r \ll r_{\rm p}$ by a factor of $1 + \Delta T/(\dot{M}r^2 \Omega_{\rm K})$, where $\dot{M} \equiv- F_{\rm M}$ ($>0$) is the net mass accretion rate and $\Delta T$ is the net torque from the planet (see their equation (19)). For $(\alpha, M_*/M_{\rm p}, h_{\rm p}) = (10^{-3}, 10^{-3}, 0.05)$, which correspond to our Run 4, \citet{Dempsey20} found $\Delta T \approx 2\dot{M}r_{\rm p}^2 \Omega_{\rm p}$, indicating a pileup factor of $\Delta T/(\dot{M}r^2 \Omega_{\rm K}) \approx 2(r/r_{\rm p})^{-1/2}$. However, no clear enhancement in the surface density across the gap is visible in Run 4, even at the end of the simulation. This is likely due to our choice of radial boundary conditions, which fix the fluid quantities at the boundaries to their initial values, and/or the computational time being insufficient for the entire disk to relax to a viscous steady state \citep{Dempsey20}.

In summary, it is possible that the gap models presented in this study are affected by the fact that our simulations have not yet reached a global viscous steady state. Before closing this subsection, we discuss how this limitation could be addressed in future work. One straightforward way to account for the relative enhancement in the surface density across the gap is to connect  the gap surface densities at $r = r_{\rm in}$ and $r = r_{\rm out}$ to $\dot{M}/(3\pi \nu)$ and $(1 + \Delta T/(\dot{M}r^2 \Omega_{\rm K}))\dot{M}/(3\pi \nu)$, respectively, as described in equation~(19) of \citet{Dempsey20}. This approach becomes feasible if $\Delta T/\dot{M}$ can be expressed as a function of disk and planetary parameters (see, e.g., equations~(29) and (32) of \citealt{Dempsey20}, although these are derived for locally isothermal disks). 
For the surface density within the gap ($r_{\rm in} < r < r_{\rm out}$), we expect that equations~\eqref{eq:Sigma_analytic_in} and \eqref{eq:Sigma_analytic_out} still serve as a good approximation in a global viscous steady state, because the radial profiles of the specific entropy jumps across the primary shocks---which are responsible for shaping the gap structure---are largely independent of the surface density structure. This was first noticed by \citet{Ono25} for disks with different initial surface density slopes (see the left panel of their figure 8). The robustness of the $\delta$ profile is also evident in figure~\ref{fig:time}, which shows that the specific entropy jumps across the primary shocks (bottom panel) relax to their final profiles significantly earlier than does the surface density (top panel). In future work, we will investigate whether such an extension of our gap model to a global viscous steady state is possible.
}

\subsection{Beyond the classical viscous disk model}
\label{sec:alpha}
Our simulations employed the classical viscous disk model with the standard $\alpha$-viscosity prescription. This choice was motivated by its simplicity: the steady-state structure of a planet-induced gap in a viscous disk is simply determined by the balance between the radially diverging flow induced by spiral shocks and the converging viscous flow (see sections~\ref{sec:bridge} and \ref{sec:semi}).

However, the fundamental question remains as to whether the simple viscous disk model is applicable to protoplanetary disks \citep[for recent reviews, see][]{Lesur23,Manara23,Pascucci23,Aikawa24}. Turbulence has classically been considered the primary source of macroscopic viscosity, but recent observations show that only a few protoplanetary disks exhibit detectable signatures of turbulence \citep[e.g.,][]{Rosotti23,Villenave25}.
Meanwhile, state-of-the-art magnetohydrodynamical (MHD) models of protoplanetary disks indicate that disk accretion is more likely driven by MHD winds than by turbulence, as the low ionization levels in cold disk regions largely suppress MHD turbulence \citep[e.g.,][]{Lesur23,Pascucci23}.
Wind-driven accretion disks are fundamentally distinct from viscous disks in that the radial gas flow is advective and directed inward, as the winds remove angular momentum from the disk \citep[see, e.g., subsection 3.2.2 of][]{Manara23}. The mechanisms governing the structure of planet-induced gaps in laminar, wind-driven MHD disks have yet to be fully understood and modeled \citep{Aoyama23,Wafflard-FernandezLesur23,Hu25,HammerLin25}.
Moreover, our empirical formula for the entropy jumps across the spiral shocks, equation~\eqref{eq:delta_formula}, is not directly applicable to such laminar disks because it explicitly depends on the viscosity parameter $\alpha$. As discussed in section~\ref{sec:scaling}, this $\alpha$ dependence is presumably linked to the surface density in the region where the primary spiral arms are excited. 
We plan to investigate the gap structure and spiral shock strengths in inviscid, wind-driven accretion disks in a forthcoming paper (Tominaga et al., in prep.).

Using an MHD-based model is also essential for achieving a more realistic understanding of the disk temperature structure. In the classical viscous disk with a spatially uniform $\alpha$ value, accretional heating predominantly occurs at the midplane, where the gas density is highest. Typically, this viscous heating dominates over heating from planet-induced shocks unless the planet is as massive as Jupiter or larger (see subsection 4.1 of \citealt{Ono25}). In contrast, in protoplanetary disks with poorly ionized interiors, heating associated with MHD-driven accretion  preferentially takes place near the disk surface, where the generated heat  radiates away efficiently \citep{Hirose11,Mori19,Mori21,Bethune20,Kondo23}. Our 2D viscous disk simulations emulate this inefficient internal heating in MHD accretion disks by switching off viscous heating arising from Keplerian shear. Our ongoing simulations using inviscid, wind-driven disk models will provide better insight into how planet-induced shock heating impacts the thermal structure of realistic protoplanetary disks.

\subsection{Need for a more realistic thermal relaxation model}
\label{sec:beta}

Another important limitation of our simulations is their use of the $\beta$-relaxation model (equation~\eqref{eq:Qrelax}) with a {\it constant} relaxation parameter $\beta$. \cite{Ono25} showed that as long as $M_{\rm p}/M_* < 10^{-3}$, $h_{\rm p} = 0.05$, $\alpha = 10^{-3}$, and $10 \leq \beta \leq 10^{2}$, the magnitudes of the entropy jumps across the primary shocks are insensitive to the choice of the $\beta$ constant,  This finding motivated us to fix $\beta$ to be $100$ in this study. Meanwhile, \cite{Ono25} also found that the planet-induced shock heating destabilizes the disk temperature structure for a higher $\beta$ value of $\beta = 10^{2.5}$ with $M_{\rm p}/M_* = 10^{-3}$, $h_{\rm p} = 0.05$, and $\alpha = 10^{-3}$ (see their section 2.4).
This instability was found to vanish when $\alpha$ was reduced to $10^{-3.5}$ (Run 11 of \citealt{Ono25}). 
Consequently, we were unable to explore parameter space with $\beta > 100$, which is more relevant to the inner {few au} of protoplanetary disks  \citep{Ziampras20}.

The thermal instability mentioned above likely arises from the weak temperature dependence of the constant-$\beta$ relaxation prescription, $Q_{\rm relax}$ (equation~\eqref{eq:Qrelax}). The first term in $Q_{\rm relax}$ cools the disk at a rate linearly proportional to both $\Sigma$ and $T$. Importantly, the same holds for the shock heating rate given by equation~\eqref{eq:Qshock}, since $c_T^2 \propto T$, and since $\delta$ is independent of $\Sigma$ and $T$ according to our {empirical} fit (equation~\eqref{eq:delta_formula}). Therefore, the ratio between the heating and cooling rates is independent of both $\Sigma$ and $T$. Once the shock heating rate exceeds the prescribed cooling rate, it continues to dominate, leading to a runaway increase in temperature. Using equations~\eqref{eq:cT2}, \eqref{eq:Qrelax}, and \eqref{eq:Qshock} together with the approximation $\Omega_{\rm init} \approx \Omega_{\rm K}$, we predict that this runaway heating occurs when the following criterion is met:
\begin{equation}
\delta \left|1-\frac{\Omega_{\rm p}}{\Omega_{\rm K}}\right| > \frac{2\pi}{\beta}.
\label{eq:thermal}
\end{equation}
Note that this prediction is consistent with the equilibrium temperature predicted by equation~\eqref{eq:T_predict} with $f_{\rm heat} = 1$, which diverges as $\delta |1-\Omega_{\rm p}/\Omega_{\rm K}| \to 2\pi/\beta$.

\begin{figure}[t]
\begin{center}
\includegraphics[width=\hsize, bb=0 0 364.1203125 258.068125]{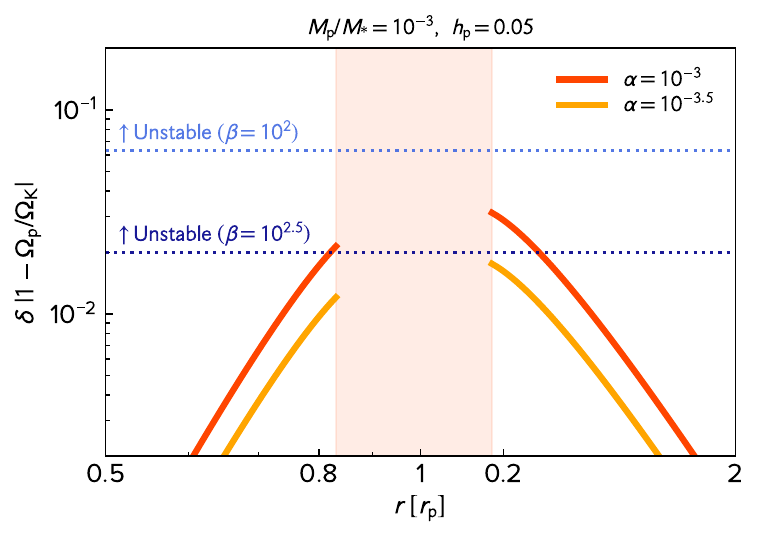}
\end{center}
\caption{Application of the thermal instability criterion (equation~\eqref{eq:thermal}) to disk models with $M_{\rm p}/M_* = 10^{-3}$ and $h_{\rm p} = 0.05$. Solid lines show $\delta |1-\Omega_{\rm p}/\Omega_{\rm K}|$ for $\alpha = 10^{-3}$ and $10^{-3.5}$, calculated using the {empirical} expression for $\delta$ given by equation~\eqref{eq:delta_formula}, while dotted lines mark $2\pi/\beta$ for $\beta = 10^{2}$ and $10^{2.5}$. Run 4 of this study corresponds to $(\alpha, \beta) =(10^{-3}, 10^2)$. Equation~\eqref{eq:thermal} predicts  runaway heating by planet-induced spiral shocks in regions where the solid line exceeds the dotted line. The {empirical} expression for $\delta$ used here is valid outside the shaded region of $|r - r_{\rm p}| < \varepsilon = 0.17r_{\rm p}$.
}
\label{fig:thermal}
\end{figure}
To test if equation \eqref{eq:thermal} explains the thermal instability observed by \citet{Ono25}, we plot in figure~\ref{fig:thermal} the left- and right-hand sides of equation~\eqref{eq:thermal} as a function of $r$ for $\alpha = 10^{-3}$ and $10^{-3.5}$, $\beta = 10^2$ and $10^{2.5}$, $M_{\rm p}/M_* = 10^{-3}$, and $h_{\rm p} = 0.05$. For $\delta$, we use the {empirical} fit given by equation~\eqref{eq:delta_formula}, which is valid at $|r - r_{\rm p}| > 0.17r_{\rm p}$. Even when this fit is extrapolated toward $r \to r_{\rm p}$, the product $\delta |1-\Omega_{\rm p}/\Omega_{\rm K}|$ does not increase further, as the factor $|1-\Omega_{\rm p}/\Omega_{\rm K}|$ suppresses the overall value. We find that for $\alpha = 10^{-3}$ and $\beta = 10^{2.5}$, the parameters under which \citet{Ono25} observed thermal instability, the predicted instability criterion is satisfied in regions close to the planet's orbit, at $r \sim 0.8r_{\rm p}$ and $1.2r_{\rm p}$. These regions vanish when $\alpha$ is reduced to $10^{-3.5}$, again consistent with the finding of \citet{Ono25}.

The above analysis implies that the thermal instability can be avoided by adopting a more realistic thermal relaxation prescription. In inner disk regions with a high optical depth $\tau \propto \Sigma$, radiative cooling from the disk surface leads to midplane cooling at a rate $\propto T^4/\tau \propto T^4/\Sigma$, where $T$ stands for the midplane temperature (see, e.g., equation~(11) of \citealt{Ziampras20}; equations~(21) and (22) of \citealt{Okuzumi22}). Comparing this with the $\beta$-cooling rate, which scales as $\Sigma T/\beta$, we find that radiative cooling is equivalent to $\beta$-cooling with a {\it variable} relaxation parameter of $\beta \propto \Sigma^2/T^{3}$. Since the cooling timescale scales with $\beta$, an increase in $T$ leads to faster radiative cooling, thereby suppressing the thermal instability that could occur under a temperature-independent $\beta$ prescription. In future work, we will test this prediction and extend our modeling to higher effective $\beta$ values using a realistic radiative cooling model.

\subsection{Importance of secondary and higher-order spiral shocks}
\label{sec:secondary}
As noted in subsection~\ref{sec:simulationresults}, a planet generally induces multiple spiral arms, each of which steepens into a shock at a different distance from the planet. Among them, the primary shocks play the dominant role in carving the gap near the planetary orbit. In contrast, the second and higher-order spiral arms tend to dominate over the primary shocks at large radii, because their associated arms retain their energy and angular momentum until they shock.
Consequently, the secondary shocks can give non-negligible contributions to the outskirts of the gap, as discussed in section~\ref{sec:analytic}. Secondary and higher-order shocks can even open distinct gaps when viscosity is low (i.e., $\alpha \ll 10^{-3}$, \citealt{Bae17,Dong17}). 

In principle, our analytic gap model can be improved by adding the entropy jumps across secondary and higher-order shocks to the radial integrals in equations~\eqref{eq:Sigma_semi_in} and \eqref{eq:Sigma_semi_out}. 
However, the radial computational domain adopted in our simulations is not wide enough to capture the full extent of the secondary shocks, making it difficult to derive a reliable {empirical} fit for their $\delta$ profiles. Moreover, unlike the primary shocks, the radial locations where secondary and higher-order shocks develop depend on the disk's sound velocity distribution, and hence on its temperature distribution, as noted in subsection~\ref{sec:simulationresults}. This adds complexity to empirical modeling of the entropy jumps across secondary and higher-order shocks.  We defer this task to future work employing a wider computational domain.

We note that secondary and higher-order spiral shocks can also play a significant role in heating disk regions distant from the planet. Indeed, the simulations by \citet{Ziampras20} indicate that shock heating by a Jupiter-mass planet remains significant away from the planetary orbit. 
This extended heating is likely due to secondary and higher-order spiral shocks, as the primary shocks typically decay substantially within the gap region. In a future study (Shimizu et al., in prep.), we will investigate the impact of secondary and higher-order spiral shocks on the disk temperature structure and the location of the snow line using simulations with a wider radial computational domain.

\section{Conclusions}\label{sec:conclusions}

We have constructed self-consistent {one-dimensional} models for {secular} heating and gap opening by planet-induced shocks {in the inner regions of protoplanetary disks}. Our models quantify the rates of shock-driven gap opening and heating in terms of the specific entropy jumps across the shocks (figure~\ref{fig:concept}). To empirically model the specific entropy jumps, we conducted 2D hydrodynamic simulations of disks with an embedded planet, adopting the standard $\alpha$ viscosity and $\beta$ relaxation prescriptions. {The presented models are most applicable to the inner $\lesssim 10~\rm au$ disk region, where the optical thickness is large and cooling is slow ($\beta \gg 1$).} Our key findings are summarized as follows.

\begin{enumerate}
\item 
The entropy jumps across the primary spiral shocks approximately follow a simple power-law function of the viscosity parameter ($\alpha$), the initial disk aspect ratio at the planet's orbit ($h_{\rm p}$), the planet's mass normalized by the thermal mass ($M_{\rm p}/M_{\rm th}$), and the distance from the central star normalized by the planet's orbital radius ($r/r_{\rm p}$) (equation~\eqref{eq:delta_formula}, the second row of figure~\ref{fig:sigma_vr}). Our {empirical} fit reproduces the entropy jumps at the primary shocks measured in our {\rm steady-state simulation results} to within $20\%$ accuracy. 

\item 
Using the theoretical relationship between angular momentum deposition and shock heating by spiral waves (equation~\eqref{eq:Lambda_shock}, \citealt{GoodmanRafikov01,Rafikov16,ArzamasskiyRafikov18}), we have formulated the azimuthally averaged radial gas velocity induced by the shocks, $v_{r,\rm shock}$, as a function of entropy jump $\delta$ (section~\ref{sec:bridge}, equation~\eqref{eq:vr_shock}). Our simulations confirm that in a steady-state gap, the predicted shock-induced approximately balances the viscosity-driven radial velocity, $v_{r,\rm visc}$, which acts to refill the gap (see the bottom row of figure~\ref{fig:sigma_vr}). 

\item 
Conversely, by assuming an exact balance between $v_{r,\rm shock}$ and $v_{r,\rm visc}$ in the gap region, we have derived expressions for the gap surface density profile as a function of the radial integral of $\delta$ (section~\ref{sec:semi}, equations~\eqref{eq:Sigma_semi_in} and \eqref{eq:Sigma_semi_out}). Since $\delta$ also characterizes shock-induced heating, the temperature distribution can be determined self-consistently, for example by balancing shock heating with thermal relaxation (equation~\eqref{eq:T_predict}). When the entropy jump distribution is taken directly from simulations, the model reproduces the  temperature and surface density profiles in regions with detectable spiral shocks to within errors of approximately 4\% and 30\%, respectively.

\item 
By employing the power-law empirical formula for the entropy jumps across the primary shocks, we have also constructed a full-analytic, self-consistent model for the gap temperature and surface density profiles (section~\ref{sec:analytic}, equations~\eqref{eq:T_predict}--\eqref{eq:Sigma_analytic_out}). Combined with an existing analytic model for the gap floor surface density \citep{KanagawaMuto15,KanagawaTanaka15}, our full-analytic model predicts both  temperature and surface density profiles across the entire gap. In terms of  surface density, our model is as accurate as or even more accurate than the widely used deep-gap model by \citet{KanagawaTanaka17} (appendix~\ref{sec:Kanagawa}, figure~\ref{fig:Kanagawa}). Incorporating the contributions from secondary shocks will further improve the accuracy of the predicted temperature and surface density profiles at the outskirts of the gap.

\end{enumerate}

The models presented in this work will pave the way for studying how a giant protoplanet influences the mass distribution and compositional gradients of solids in the {inner $\lesssim$ 10 au disk region}. {We note, however, that several caveats remain and should be addressed in future work. First, although the gap structures observed in our hydrodynamical simulations appear to be in a nearly steady state, the simulated disks as a whole have not reached a global viscous steady state characterized by a radially constant net accretion flux (subsection~\ref{sec:steady}). Some modifications will be necessary to extend our models to such a globally steady-state configuration. Furthermore,} our models have so far only been tested against the classical $\alpha$-disk model with a standard $\beta$-relaxation prescription. Extending it to laminar, MHD-driven accretion disk models is the subject of future work (subsection~\ref{sec:alpha}). Because of the thermal instability arising from the constant-$\beta$ model, our simulations have been restricted to a moderately high value of $\beta = 100$. Simulations using a more realistic thermal relaxation model and a wider radial computational domain are necessary to extend our models to the inner {$\sim $ 1 au} of protoplanetary disks, where significantly higher $\beta$ values are expected and secondary and higher-order spiral shocks may contribute substantially to the disk temperature structure (subsections~\ref{sec:beta} and~\ref{sec:secondary}).

\begin{ack}
We thank Alex Ziampras, Jaehan Bae, Hidekazu Tanaka, Shota Sato, and Ayumu Kuwahara for stimulating discussions, and Tomohiro Ono for providing us with the simulation code, {and an anonymous reviewer for constructive and insightful comments on limitations of the presented models.}
\end{ack}

\section*{Funding}
This work was supported by JSPS KAKENHI (Grant Numbers JP19K03926, JP20H00182, JP20H01948, JP23H01227, and JP23K25923) and NINS Astrobiology Center program research
(Grant Number AB0705).

\bibliographystyle{apj}
\bibliography{SurfaceAccretion}

@PREAMBLE{"\def\SortNoop#1{}"}

@PREAMBLE{"\DeclareAbbreviation\icarus{Icarus}"}

@PREAMBLE{"\DeclareAbbreviation\nar{NewAR}"}

@ARTICLE{Cimerman24,
       author = {{Cimerman}, Nicolas P. and {Rafikov}, Roman R. and {Miranda}, Ryan},
        title = "{Torque wiggles - a robust feature of the global disc-planet interaction}",
      journal = {\mnras},
         year = 2024,
        month = mar,
       volume = {529},
       number = {1},
        pages = {425-443},
          doi = {10.1093/mnras/stae467},
archivePrefix = {arXiv},
       eprint = {2306.07341},
 primaryClass = {astro-ph.EP},
       adsurl = {https://ui.adsabs.harvard.edu/abs/2024MNRAS.529..425C}
}

@ARTICLE{Miranda19a,
       author = {{Miranda}, Ryan and {Rafikov}, Roman R.},
        title = "{Multiple Spiral Arms in Protoplanetary Disks: Linear Theory}",
      journal = {\apj},
         year = 2019,
        month = apr,
       volume = {875},
       number = {1},
          eid = {37},
        pages = {37},
          doi = {10.3847/1538-4357/ab0f9e},
archivePrefix = {arXiv},
       eprint = {1811.09628},
 primaryClass = {astro-ph.EP},
       adsurl = {https://ui.adsabs.harvard.edu/abs/2019ApJ...875...37M}
}

@ARTICLE{Kley99,
       author = {{Kley}, W.},
        title = "{Mass flow and accretion through gaps in accretion discs}",
      journal = {\mnras},
         year = 1999,
        month = mar,
       volume = {303},
       number = {4},
        pages = {696-710},
          doi = {10.1046/j.1365-8711.1999.02198.x},
archivePrefix = {arXiv},
       eprint = {astro-ph/9809253},
 primaryClass = {astro-ph},
       adsurl = {https://ui.adsabs.harvard.edu/abs/1999MNRAS.303..696K}
}

@ARTICLE{Juhasz15,
       author = {{Juh{\'a}sz}, A. and {Benisty}, M. and {Pohl}, A. and {Dullemond}, C.~P. and {Dominik}, C. and {Paardekooper}, S. -J.},
        title = "{Spiral arms in scattered light images of protoplanetary discs: are they the signposts of planets?}",
      journal = {\mnras},
         year = 2015,
        month = aug,
       volume = {451},
       number = {2},
        pages = {1147-1157},
          doi = {10.1093/mnras/stv1045},
archivePrefix = {arXiv},
       eprint = {1412.3412},
 primaryClass = {astro-ph.EP},
       adsurl = {https://ui.adsabs.harvard.edu/abs/2015MNRAS.451.1147J}
}

@ARTICLE{FungDong15,
       author = {{Fung}, Jeffrey and {Dong}, Ruobing},
        title = "{Inferring Planet Mass from Spiral Structures in Protoplanetary Disks}",
      journal = {\apjl},
         year = 2015,
        month = dec,
       volume = {815},
       number = {2},
          eid = {L21},
        pages = {L21},
          doi = {10.1088/2041-8205/815/2/L21},
archivePrefix = {arXiv},
       eprint = {1511.01178},
 primaryClass = {astro-ph.EP},
       adsurl = {https://ui.adsabs.harvard.edu/abs/2015ApJ...815L..21F}
}

@ARTICLE{ZhuDong15,
       author = {{Zhu}, Zhaohuan and {Dong}, Ruobing and {Stone}, James M. and {Rafikov}, Roman R.},
        title = "{The Structure of Spiral Shocks Excited by Planetary-mass Companions}",
      journal = {\apj},
         year = 2015,
        month = nov,
       volume = {813},
       number = {2},
          eid = {88},
        pages = {88},
          doi = {10.1088/0004-637X/813/2/88},
archivePrefix = {arXiv},
       eprint = {1507.03599},
 primaryClass = {astro-ph.SR},
       adsurl = {https://ui.adsabs.harvard.edu/abs/2015ApJ...813...88Z}
}

@article{Rafikov16,
archivePrefix = {arXiv},
arxivId = {1601.03009},
author = {Rafikov, Roman R.},
doi = {10.3847/0004-637X/831/2/122},
eprint = {1601.03009},
issn = {1538-4357},
journal = {\apj},
month = {nov},
number = {2},
pages = {122},
title = {{PROTOPLANETARY DISK HEATING AND EVOLUTION DRIVEN BY SPIRAL DENSITY WAVES}},
url = {https://iopscience.iop.org/article/10.3847/0004-637X/831/2/122},
volume = {831},
year = {2016}
}

@ARTICLE{Bae18a,
       author = {{Bae}, Jaehan and {Zhu}, Zhaohuan},
        title = "{Planet-driven Spiral Arms in Protoplanetary Disks. I. Formation Mechanism}",
      journal = {\apj},
         year = 2018,
        month = jun,
       volume = {859},
       number = {2},
          eid = {118},
        pages = {118},
          doi = {10.3847/1538-4357/aabf8c},
archivePrefix = {arXiv},
       eprint = {1711.08161},
 primaryClass = {astro-ph.EP},
       adsurl = {https://ui.adsabs.harvard.edu/abs/2018ApJ...859..118B}
}

@ARTICLE{Bae18b,
       author = {{Bae}, Jaehan and {Zhu}, Zhaohuan},
        title = "{Planet-driven Spiral Arms in Protoplanetary Disks. II. Implications}",
      journal = {\apj},
         year = 2018,
        month = jun,
       volume = {859},
       number = {2},
          eid = {119},
        pages = {119},
          doi = {10.3847/1538-4357/aabf93},
archivePrefix = {arXiv},
       eprint = {1711.08166},
 primaryClass = {astro-ph.EP},
       adsurl = {https://ui.adsabs.harvard.edu/abs/2018ApJ...859..119B}
}

@ARTICLE{Ono25,
       author = {{Ono}, Tomohiro and {Okamura}, Tatsuki and {Okuzumi}, Satoshi and {Muto}, Takayuki},
        title = "{Modeling protoplanetary disk heating by planet-induced spiral shocks}",
      journal = {\pasj},
         year = 2025,
        month = feb,
       volume = {77},
       number = {1},
        pages = {149-161},
          doi = {10.1093/pasj/psae106},
archivePrefix = {arXiv},
       eprint = {2411.09940},
 primaryClass = {astro-ph.EP},
       adsurl = {https://ui.adsabs.harvard.edu/abs/2025PASJ...77..149O}
}

@ARTICLE{GoodmanRafikov01,
       author = {{Goodman}, J. and {Rafikov}, R.~R.},
        title = "{Planetary Torques as the Viscosity of Protoplanetary Disks}",
      journal = {\apj},
         year = 2001,
        month = may,
       volume = {552},
       number = {2},
        pages = {793-802},
          doi = {10.1086/320572},
archivePrefix = {arXiv},
       eprint = {astro-ph/0010576},
 primaryClass = {astro-ph},
       adsurl = {https://ui.adsabs.harvard.edu/abs/2001ApJ...552..793G}
}

@ARTICLE{Rafikov02a,
       author = {{Rafikov}, R.~R.},
        title = "{Nonlinear Propagation of Planet-generated Tidal Waves}",
      journal = {\apj},
         year = 2002,
        month = apr,
       volume = {569},
       number = {2},
        pages = {997-1008},
          doi = {10.1086/339399},
archivePrefix = {arXiv},
       eprint = {astro-ph/0110496},
 primaryClass = {astro-ph},
       adsurl = {https://ui.adsabs.harvard.edu/abs/2002ApJ...569..997R}
}

@ARTICLE{Rafikov02b,
       author = {{Rafikov}, R.~R.},
        title = "{Planet Migration and Gap Formation by Tidally Induced Shocks}",
      journal = {\apj},
         year = 2002,
        month = jun,
       volume = {572},
       number = {1},
        pages = {566-579},
          doi = {10.1086/340228},
archivePrefix = {arXiv},
       eprint = {astro-ph/0110540},
 primaryClass = {astro-ph},
       adsurl = {https://ui.adsabs.harvard.edu/abs/2002ApJ...572..566R}
}

@article{Ziampras20,
       author = {{Ziampras}, Alexandros and {Ataiee}, Sareh and {Kley}, Wilhelm and {Dullemond}, Cornelis P. and {Baruteau}, Cl{\'e}ment},
        title = "{The impact of planet wakes on the location and shape of the water ice line in a protoplanetary disk}",
      journal = {\aap},
         year = 2020,
        month = jan,
       volume = {633},
          eid = {A29},
        pages = {A29},
          doi = {10.1051/0004-6361/201936495},
archivePrefix = {arXiv},
       eprint = {1910.08560},
 primaryClass = {astro-ph.EP},
       adsurl = {https://ui.adsabs.harvard.edu/abs/2020A&A...633A..29Z}
}

@ARTICLE{Rosotti23,
       author = {{Rosotti}, Giovanni P.},
        title = "{Empirical constraints on turbulence in proto-planetary discs}",
      journal = {\nar},
         year = 2023,
        month = jun,
       volume = {96},
          eid = {101674},
        pages = {101674},
          doi = {10.1016/j.newar.2023.101674},
archivePrefix = {arXiv},
       eprint = {2302.01433},
 primaryClass = {astro-ph.EP},
       adsurl = {https://ui.adsabs.harvard.edu/abs/2023NewAR..9601674R}
}

@ARTICLE{Kondo23,
       author = {{Kondo}, Katsushi and {Okuzumi}, Satoshi and {Mori}, Shoji},
        title = "{The Roles of Dust Growth in the Temperature Evolution and Snow Line Migration in Magnetically Accreting Protoplanetary Disks}",
      journal = {\apj},
         year = 2023,
        month = jun,
       volume = {949},
       number = {2},
          eid = {119},
        pages = {119},
          doi = {10.3847/1538-4357/acc840},
archivePrefix = {arXiv},
       eprint = {2205.13511},
 primaryClass = {astro-ph.EP},
       adsurl = {https://ui.adsabs.harvard.edu/abs/2023ApJ...949..119K}
}

@ARTICLE{Okuzumi22,
       author = {{Okuzumi}, Satoshi and {Ueda}, Takahiro and {Turner}, Neal J.},
        title = "{A global two-layer radiative transfer model for axisymmetric, shadowed protoplanetary disks}",
      journal = {\pasj},
         year = 2022,
        month = aug,
       volume = {74},
       number = {4},
        pages = {828-850},
          doi = {10.1093/pasj/psac040},
archivePrefix = {arXiv},
       eprint = {2201.09241},
 primaryClass = {astro-ph.EP},
       adsurl = {https://ui.adsabs.harvard.edu/abs/2022PASJ...74..828O}
}

@ARTICLE{Dullemond18,
       author = {{Dullemond}, Cornelis P. and {Birnstiel}, Tilman and {Huang}, Jane and {Kurtovic}, Nicol{\'a}s T. and {Andrews}, Sean M. and {Guzm{\'a}n}, Viviana V. and {P{\'e}rez}, Laura M. and {Isella}, Andrea and {Zhu}, Zhaohuan and {Benisty}, Myriam and {Wilner}, David J. and {Bai}, Xue-Ning and {Carpenter}, John M. and {Zhang}, Shangjia and {Ricci}, Luca},
        title = "{The Disk Substructures at High Angular Resolution Project (DSHARP). VI. Dust Trapping in Thin-ringed Protoplanetary Disks}",
      journal = {\apjl},
         year = 2018,
        month = dec,
       volume = {869},
       number = {2},
          eid = {L46},
        pages = {L46},
          doi = {10.3847/2041-8213/aaf742},
archivePrefix = {arXiv},
       eprint = {1812.04044},
 primaryClass = {astro-ph.EP},
       adsurl = {https://ui.adsabs.harvard.edu/abs/2018ApJ...869L..46D}
}

@ARTICLE{Aoyama23,
       author = {{Aoyama}, Yuhiko and {Bai}, Xue-Ning},
        title = "{Three-dimensional Global Simulations of Type-II Planet-Disk Interaction with a Magnetized Disk Wind. I. Magnetic Flux Concentration and Gap Properties}",
      journal = {\apj},
         year = 2023,
        month = mar,
       volume = {946},
       number = {1},
          eid = {5},
        pages = {5},
          doi = {10.3847/1538-4357/acb81f},
archivePrefix = {arXiv},
       eprint = {2302.01514},
 primaryClass = {astro-ph.EP},
       adsurl = {https://ui.adsabs.harvard.edu/abs/2023ApJ...946....5A}
}

@ARTICLE{Villenave25,
       author = {{Villenave}, Marion and {Rosotti}, Giovanni P. and {Lambrechts}, Michiel and {Ziampras}, Alexandros and {Pinte}, Christophe and {M{\'e}nard}, Fran{\c{c}}ois and {Stapelfeldt}, Karl R. and {Duch{\^e}ne}, Gaspard and {Baylock}, Emily and {Doi}, Kiyoaki},
        title = "{Turbulence in protoplanetary disks: A systematic analysis of dust settling in 33 disks}",
      journal = {\aap},
         year = 2025,
        month = may,
       volume = {697},
          eid = {A64},
        pages = {A64},
          doi = {10.1051/0004-6361/202553822},
archivePrefix = {arXiv},
       eprint = {2503.05872},
 primaryClass = {astro-ph.SR},
       adsurl = {https://ui.adsabs.harvard.edu/abs/2025A&A...697A..64V}
}

@ARTICLE{Miranda20a,
       author = {{Miranda}, Ryan and {Rafikov}, Roman R.},
        title = "{Planet-Disk Interaction in Disks with Cooling: Basic Theory}",
      journal = {\apj},
         year = 2020,
        month = mar,
       volume = {892},
       number = {1},
          eid = {65},
        pages = {65},
          doi = {10.3847/1538-4357/ab791a},
archivePrefix = {arXiv},
       eprint = {1911.01428},
 primaryClass = {astro-ph.EP},
       adsurl = {https://ui.adsabs.harvard.edu/abs/2020ApJ...892...65M}
}

@ARTICLE{Miranda20b,
       author = {{Miranda}, Ryan and {Rafikov}, Roman R.},
        title = "{Gaps and Rings in Protoplanetary Disks with Realistic Thermodynamics: The Critical Role of In-plane Radiation Transport}",
      journal = {\apj},
         year = 2020,
        month = dec,
       volume = {904},
       number = {2},
          eid = {121},
        pages = {121},
          doi = {10.3847/1538-4357/abbee7},
archivePrefix = {arXiv},
       eprint = {2007.13766},
 primaryClass = {astro-ph.EP},
       adsurl = {https://ui.adsabs.harvard.edu/abs/2020ApJ...904..121M}
}

@ARTICLE{ZhangZhu20,
       author = {{Zhang}, Shangjia and {Zhu}, Zhaohuan},
        title = "{The effects of disc self-gravity and radiative cooling on the formation of gaps and spirals by young planets}",
      journal = {\mnras},
         year = 2020,
        month = apr,
       volume = {493},
       number = {2},
        pages = {2287-2305},
          doi = {10.1093/mnras/staa404},
archivePrefix = {arXiv},
       eprint = {1911.01530},
 primaryClass = {astro-ph.EP},
       adsurl = {https://ui.adsabs.harvard.edu/abs/2020MNRAS.493.2287Z}
}

@ARTICLE{Muller12,
       author = {{M{\"u}ller}, T.~W.~A. and {Kley}, W. and {Meru}, F.},
        title = "{Treating gravity in thin-disk simulations}",
      journal = {\aap},
         year = 2012,
        month = may,
       volume = {541},
          eid = {A123},
        pages = {A123},
          doi = {10.1051/0004-6361/201118737},
archivePrefix = {arXiv},
       eprint = {1203.1413},
 primaryClass = {astro-ph.EP},
       adsurl = {https://ui.adsabs.harvard.edu/abs/2012A&A...541A.123M}
}

@ARTICLE{Dempsey20,
       author = {{Dempsey}, Adam M. and {Lee}, Wing-Kit and {Lithwick}, Yoram},
        title = "{Pileups and Migration Rates for Planets in Low-mass Disks}",
      journal = {\apj},
         year = 2020,
        month = mar,
       volume = {891},
       number = {2},
          eid = {108},
        pages = {108},
          doi = {10.3847/1538-4357/ab723c},
archivePrefix = {arXiv},
       eprint = {1908.02326},
 primaryClass = {astro-ph.EP},
       adsurl = {https://ui.adsabs.harvard.edu/abs/2020ApJ...891..108D}
}

@article{DeValBorro06,
   author = {M. De Val-Borro and R. G. Edgar and P. Artymowicz and P. Ciecielag and P. Cresswell and G. D'Angelo and E. J. Delgado-Donate and G. Dirksen and S. Fromang and A. Gawryszczak and H. Klahr and W. Kley and W. Lyra and F. Masset and G. Mellema and R. P. Nelson and S.- J. Paardekooper and A. Peplinski and A. Pierens and T. Plewa and K. Rice and C. Schafer and R. Speith},
   doi = {10.1111/j.1365-2966.2006.10488.x},
   issn = {0035-8711},
   issue = {2},
   journal = {\mnras},
   month = {8},
   pages = {529-558},
   title = {A comparative study of disc-planet interaction},
   volume = {370},
   url = {https://academic.oup.com/mnras/article-lookup/doi/10.1111/j.1365-2966.2006.10488.x},
   year = {2006},
}

@ARTICLE{Ziampras20b,
       author = {{Ziampras}, Alexandros and {Kley}, Wilhelm and {Dullemond}, Cornelis P.},
        title = "{Importance of radiative effects in gap opening by planets in protoplanetary disks}",
      journal = {\aap},
         year = 2020,
        month = may,
       volume = {637},
          eid = {A50},
        pages = {A50},
          doi = {10.1051/0004-6361/201937048},
archivePrefix = {arXiv},
       eprint = {2003.02298},
 primaryClass = {astro-ph.EP},
       adsurl = {https://ui.adsabs.harvard.edu/abs/2020A&A...637A..50Z}
}

@INCOLLECTION{Aikawa24,
       author = {{Aikawa}, Yuri and {Okuzumi}, Satoshi and {Pontoppidan}, Klaus},
        title = "{The Physical and Chemical Processes in Protoplanetary Disks: Constraints on the Composition of Comets}",
    booktitle = {Comets III},
         year = 2024,
       editor = {{Meech}, Karen. J. and {Combi}, Michael. R. and {Bockel{\'e}e-Morvan}, Dominique and {Raymond}, Sean. N. and {Zolensky}, Michael. E.},
        pages = {33-62},
       adsurl = {https://ui.adsabs.harvard.edu/abs/2024come.book...33A}
}

@INPROCEEDINGS{Pascucci23,
       author = {{Pascucci}, I. and {Cabrit}, S. and {Edwards}, S. and {Gorti}, U. and {Gressel}, O. and {Suzuki}, T.~K.},
        title = "{The Role of Disk Winds in the Evolution and Dispersal of Protoplanetary Disks}",
    booktitle = {Protostars and Planets VII},
         year = 2023,
       editor = {{Inutsuka}, S. and {Aikawa}, Y. and {Muto}, T. and {Tomida}, K. and {Tamura}, M.},
       series = {Astronomical Society of the Pacific Conference Series},
       volume = {534},
        month = jul,
        pages = {567},
          doi = {10.48550/arXiv.2203.10068},
archivePrefix = {arXiv},
       eprint = {2203.10068},
 primaryClass = {astro-ph.EP},
       adsurl = {https://ui.adsabs.harvard.edu/abs/2023ASPC..534..567P}
}

@INPROCEEDINGS{Manara23,
       author = {{Manara}, C.~F. and {Ansdell}, M. and {Rosotti}, G.~P. and {Hughes}, A.~M. and {Armitage}, P.~J. and {Lodato}, G. and {Williams}, J.~P.},
        title = "{Demographics of Young Stars and their Protoplanetary Disks: Lessons Learned on Disk Evolution and its Connection to Planet Formation}",
    booktitle = {Protostars and Planets VII},
         year = 2023,
       editor = {{Inutsuka}, S. and {Aikawa}, Y. and {Muto}, T. and {Tomida}, K. and {Tamura}, M.},
       series = {Astronomical Society of the Pacific Conference Series},
       volume = {534},
        month = jul,
        pages = {539},
          doi = {10.48550/arXiv.2203.09930},
archivePrefix = {arXiv},
       eprint = {2203.09930},
 primaryClass = {astro-ph.SR},
       adsurl = {https://ui.adsabs.harvard.edu/abs/2023ASPC..534..539M}
}

@INPROCEEDINGS{Lesur23,
       author = {{Lesur}, G. and {Flock}, M. and {Ercolano}, B. and {Lin}, M. -K. and {Yang}, C. and {Barranco}, J.~A. and {Benitez-Llambay}, P. and {Goodman}, J. and {Johansen}, A. and {Klahr}, H. and {Laibe}, G. and {Lyra}, W. and {Marcus}, P.~S. and {Nelson}, R.~P. and {Squire}, J. and {Simon}, J.~B. and {Turner}, N.~J. and {Umurhan}, O.~M. and {Youdin}, A.~N.},
        title = "{Hydro-, Magnetohydro-, and Dust-Gas Dynamics of Protoplanetary Disks}",
    booktitle = {Protostars and Planets VII},
         year = 2023,
       editor = {{Inutsuka}, S. and {Aikawa}, Y. and {Muto}, T. and {Tomida}, K. and {Tamura}, M.},
       series = {Astronomical Society of the Pacific Conference Series},
       volume = {534},
        month = jul,
        pages = {465},
          doi = {10.48550/arXiv.2203.09821},
archivePrefix = {arXiv},
       eprint = {2203.09821},
 primaryClass = {astro-ph.EP},
       adsurl = {https://ui.adsabs.harvard.edu/abs/2023ASPC..534..465L}
}

@ARTICLE{FungShi14,
       author = {{Fung}, Jeffrey and {Shi}, Ji-Ming and {Chiang}, Eugene},
        title = "{How Empty are Disk Gaps Opened by Giant Planets?}",
      journal = {\apj},
         year = 2014,
        month = feb,
       volume = {782},
       number = {2},
          eid = {88},
        pages = {88},
          doi = {10.1088/0004-637X/782/2/88},
archivePrefix = {arXiv},
       eprint = {1310.0156},
 primaryClass = {astro-ph.EP},
       adsurl = {https://ui.adsabs.harvard.edu/abs/2014ApJ...782...88F}
}

@ARTICLE{Bethune20,
       author = {{B{\'e}thune}, William and {Latter}, Henrik},
        title = "{Electric heating and angular momentum transport in laminar models of protoplanetary discs}",
      journal = {\mnras},
         year = 2020,
        month = jun,
       volume = {494},
       number = {4},
        pages = {6103-6119},
          doi = {10.1093/mnras/staa908},
archivePrefix = {arXiv},
       eprint = {2003.13263},
 primaryClass = {astro-ph.EP},
       adsurl = {https://ui.adsabs.harvard.edu/abs/2020MNRAS.494.6103B}
}

@ARTICLE{Mori21,
       author = {{Mori}, Shoji and {Okuzumi}, Satoshi and {Kunitomo}, Masanobu and {Bai}, Xue-Ning},
        title = "{Evolution of the Water Snow Line in Magnetically Accreting Protoplanetary Disks}",
      journal = {\apj},
         year = 2021,
        month = aug,
       volume = {916},
       number = {2},
          eid = {72},
        pages = {72},
          doi = {10.3847/1538-4357/ac06a9},
archivePrefix = {arXiv},
       eprint = {2105.13101},
 primaryClass = {astro-ph.EP},
       adsurl = {https://ui.adsabs.harvard.edu/abs/2021ApJ...916...72M}
}

@ARTICLE{Richert15,
       author = {{Richert}, Alexander J.~W. and {Lyra}, Wladimir and {Boley}, Aaron and {Mac Low}, Mordecai-Mark and {Turner}, Neal},
        title = "{On Shocks Driven by High-mass Planets in Radiatively Inefficient Disks. I. Two-dimensional Global Disk Simulations}",
      journal = {\apj},
         year = 2015,
        month = may,
       volume = {804},
       number = {2},
          eid = {95},
        pages = {95},
          doi = {10.1088/0004-637X/804/2/95},
archivePrefix = {arXiv},
       eprint = {1504.00066},
 primaryClass = {astro-ph.EP},
       adsurl = {https://ui.adsabs.harvard.edu/abs/2015ApJ...804...95R}
}

@INPROCEEDINGS{Paardekooper23,
       author = {{Paardekooper}, S. and {Dong}, R. and {Duffell}, P. and {Fung}, J. and {Masset}, F.~S. and {Ogilvie}, G. and {Tanaka}, H.},
        title = "{Planet-Disk Interactions and Orbital Evolution}",
    booktitle = {Protostars and Planets VII},
         year = 2023,
       editor = {{Inutsuka}, S. and {Aikawa}, Y. and {Muto}, T. and {Tomida}, K. and {Tamura}, M.},
       series = {Astronomical Society of the Pacific Conference Series},
       volume = {534},
        month = jul,
        pages = {685},
          doi = {10.48550/arXiv.2203.09595},
archivePrefix = {arXiv},
       eprint = {2203.09595},
 primaryClass = {astro-ph.EP},
       adsurl = {https://ui.adsabs.harvard.edu/abs/2023ASPC..534..685P}
}

@ARTICLE{Duffell13,
       author = {{Duffell}, Paul C. and {MacFadyen}, Andrew I.},
        title = "{Gap Opening by Extremely Low-mass Planets in a Viscous Disk}",
      journal = {\apj},
         year = 2013,
        month = may,
       volume = {769},
       number = {1},
          eid = {41},
        pages = {41},
          doi = {10.1088/0004-637X/769/1/41},
archivePrefix = {arXiv},
       eprint = {1302.1934},
 primaryClass = {astro-ph.EP},
       adsurl = {https://ui.adsabs.harvard.edu/abs/2013ApJ...769...41D}
}

@ARTICLE{Lyra16,
       author = {{Lyra}, Wladimir and {Richert}, Alexander J.~W. and {Boley}, Aaron and {Turner}, Neal and {Mac Low}, Mordecai-Mark and {Okuzumi}, Satoshi and {Flock}, Mario},
        title = "{On Shocks Driven by High-mass Planets in Radiatively Inefficient Disks. II. Three-dimensional Global Disk Simulations}",
      journal = {\apj},
         year = 2016,
        month = feb,
       volume = {817},
       number = {2},
          eid = {102},
        pages = {102},
          doi = {10.3847/0004-637X/817/2/102},
archivePrefix = {arXiv},
       eprint = {1511.02988},
 primaryClass = {astro-ph.EP},
       adsurl = {https://ui.adsabs.harvard.edu/abs/2016ApJ...817..102L}
}

@ARTICLE{Mori19,
   author = {{Mori}, S. and {Bai}, X.-N. and {Okuzumi}, S.},
    title = "{Temperature Structure in the Inner Regions of Protoplanetary Disks: Inefficient Accretion Heating Controlled by Nonideal Magnetohydrodynamics}",
  journal = {\apj},
archivePrefix = "arXiv",
   eprint = {1901.06921},
 primaryClass = "astro-ph.EP",
     year = 2019,
    month = feb,
   volume = 872,
      eid = {98},
    pages = {98},
      doi = {10.3847/1538-4357/ab0022},
   adsurl = {http://adsabs.harvard.edu/abs/2019ApJ...872...98M}
}

@article{Stone20,
	doi = {10.3847/1538-4365/ab929b},
	url = {https://doi.org/10.3847/1538-4365/ab929b},
	year = 2020,
	month = {jun},
	publisher = {American Astronomical Society},
	volume = {249},
	number = {1},
	pages = {4},
	author = {James M. Stone and Kengo Tomida and Christopher J. White and Kyle G. Felker},
	title = {The Athena$\mathplus$ $\mathplus$ Adaptive Mesh Refinement Framework: Design and Magnetohydrodynamic Solvers},
	journal = {\apjs},
}

@ARTICLE{Kruijer20,
       author = {{Kruijer}, Thomas S. and {Kleine}, Thorsten and {Borg}, Lars E.},
        title = "{The great isotopic dichotomy of the early Solar System}",
      journal = {Nature Astronomy},
         year = 2020,
        month = jan,
       volume = {4},
        pages = {32-40},
          doi = {10.1038/s41550-019-0959-9},
       adsurl = {https://ui.adsabs.harvard.edu/abs/2020NatAs...4...32K}
}

@ARTICLE{Homma24,
       author = {{Homma}, Kazuaki A. and {Okuzumi}, Satoshi and {Arakawa}, Sota and {Fukai}, Ryota},
        title = "{Isotopic variation of non-carbonaceous meteorites caused by dust leakage across the Jovian gap in the solar nebula}",
      journal = {\pasj},
         year = 2024,
        month = oct,
       volume = {76},
       number = {5},
        pages = {881-894},
          doi = {10.1093/pasj/psae052},
archivePrefix = {arXiv},
       eprint = {2405.20553},
 primaryClass = {astro-ph.EP},
       adsurl = {https://ui.adsabs.harvard.edu/abs/2024PASJ...76..881H}
}

@ARTICLE{Goldreich79,
       author = {{Goldreich}, P. and {Tremaine}, S.},
        title = "{The excitation of density waves at the Lindblad and corotation resonances by an external potential.}",
      journal = {\apj},
         year = 1979,
        month = nov,
       volume = {233},
        pages = {857-871},
          doi = {10.1086/157448},
       adsurl = {https://ui.adsabs.harvard.edu/abs/1979ApJ...233..857G}
}

@ARTICLE{Duffell15,
       author = {{Duffell}, Paul C.},
        title = "{A Simple Analytical Model for Gaps in Protoplanetary Disks}",
      journal = {\apjl},
         year = 2015,
        month = jul,
       volume = {807},
       number = {1},
          eid = {L11},
        pages = {L11},
          doi = {10.1088/2041-8205/807/1/L11},
archivePrefix = {arXiv},
       eprint = {1505.03514},
 primaryClass = {astro-ph.EP},
       adsurl = {https://ui.adsabs.harvard.edu/abs/2015ApJ...807L..11D}
}

@ARTICLE{Kleine20,
       author = {{Kleine}, T. and {Budde}, G. and {Burkhardt}, C. and {Kruijer}, T.~S. and {Worsham}, E.~A. and {Morbidelli}, A. and {Nimmo}, F.},
        title = "{The Non-carbonaceous-Carbonaceous Meteorite Dichotomy}",
      journal = {\ssr},
         year = 2020,
        month = may,
       volume = {216},
       number = {4},
          eid = {55},
        pages = {55},
          doi = {10.1007/s11214-020-00675-w},
       adsurl = {https://ui.adsabs.harvard.edu/abs/2020SSRv..216...55K}
}

@ARTICLE{Mizuno80,
       author = {{Mizuno}, H.},
        title = "{Formation of the Giant Planets}",
      journal = {Progress of Theoretical Physics},
         year = 1980,
        month = aug,
       volume = {64},
       number = {2},
        pages = {544-557},
          doi = {10.1143/PTP.64.544},
       adsurl = {https://ui.adsabs.harvard.edu/abs/1980PThPh..64..544M}
}

@ARTICLE{Pollack96,
       author = {{Pollack}, James B. and {Hubickyj}, Olenka and {Bodenheimer}, Peter and {Lissauer}, Jack J. and {Podolak}, Morris and {Greenzweig}, Yuval},
        title = "{Formation of the Giant Planets by Concurrent Accretion of Solids and Gas}",
      journal = {\icarus},
         year = 1996,
        month = nov,
       volume = {124},
       number = {1},
        pages = {62-85},
          doi = {10.1006/icar.1996.0190},
       adsurl = {https://ui.adsabs.harvard.edu/abs/1996Icar..124...62P}
}

@ARTICLE{Hu25,
       author = {{Hu}, Xiao and {Li}, Zhi-Yun and {Bae}, Jaehan and {Zhu}, Zhaohuan},
        title = "{3D gap opening in non-ideal MHD protoplanetary discs: asymmetric accretion, meridional vortices, and observational signatures}",
      journal = {\mnras},
         year = 2025,
        month = jan,
       volume = {536},
       number = {2},
        pages = {1374-1388},
          doi = {10.1093/mnras/stae2681},
archivePrefix = {arXiv},
       eprint = {2403.18292},
 primaryClass = {astro-ph.EP},
       adsurl = {https://ui.adsabs.harvard.edu/abs/2025MNRAS.536.1374H}
}

@ARTICLE{HammerLin25,
       author = {{Hammer}, Michael and {Lin}, Min-Kai},
        title = "{An MHD-based model for wind-driven disc{\textendash}planet interactions}",
      journal = {\mnras},
         year = 2025,
        month = jun,
       volume = {540},
       number = {2},
        pages = {1507-1526},
          doi = {10.1093/mnras/staf805},
archivePrefix = {arXiv},
       eprint = {2505.08505},
 primaryClass = {astro-ph.EP},
       adsurl = {https://ui.adsabs.harvard.edu/abs/2025MNRAS.540.1507H}
}

@ARTICLE{Wafflard-FernandezLesur23,
       author = {{Wafflard-Fernandez}, Gaylor and {Lesur}, Geoffroy},
        title = "{Planet-disk-wind interaction: The magnetized fate of protoplanets}",
      journal = {\aap},
         year = 2023,
        month = sep,
       volume = {677},
          eid = {A70},
        pages = {A70},
          doi = {10.1051/0004-6361/202245305},
archivePrefix = {arXiv},
       eprint = {2305.11784},
 primaryClass = {astro-ph.SR},
       adsurl = {https://ui.adsabs.harvard.edu/abs/2023A&A...677A..70W}
}

@ARTICLE{Shibaike20,
       author = {{Shibaike}, Y. and {Alibert}, Y.},
        title = "{Planetesimal formation at the gas pressure bump following a migrating planet. I. Basic characteristics of the new formation model}",
      journal = {\aap},
         year = 2020,
        month = dec,
       volume = {644},
          eid = {A81},
        pages = {A81},
          doi = {10.1051/0004-6361/202039086},
archivePrefix = {arXiv},
       eprint = {2010.10594},
 primaryClass = {astro-ph.EP},
       adsurl = {https://ui.adsabs.harvard.edu/abs/2020A&A...644A..81S}
}

@ARTICLE{Dong17,
       author = {{Dong}, Ruobing and {Li}, Shengtai and {Chiang}, Eugene and {Li}, Hui},
        title = "{Multiple Disk Gaps and Rings Generated by a Single Super-Earth}",
      journal = {\apj},
         year = 2017,
        month = jul,
       volume = {843},
       number = {2},
          eid = {127},
        pages = {127},
          doi = {10.3847/1538-4357/aa72f2},
archivePrefix = {arXiv},
       eprint = {1705.04687},
 primaryClass = {astro-ph.EP},
       adsurl = {https://ui.adsabs.harvard.edu/abs/2017ApJ...843..127D}
}

@ARTICLE{Bae17,
       author = {{Bae}, Jaehan and {Zhu}, Zhaohuan and {Hartmann}, Lee},
        title = "{On the Formation of Multiple Concentric Rings and Gaps in Protoplanetary Disks}",
      journal = {\apj},
         year = 2017,
        month = dec,
       volume = {850},
       number = {2},
          eid = {201},
        pages = {201},
          doi = {10.3847/1538-4357/aa9705},
archivePrefix = {arXiv},
       eprint = {1706.03066},
 primaryClass = {astro-ph.EP},
       adsurl = {https://ui.adsabs.harvard.edu/abs/2017ApJ...850..201B}
}

@ARTICLE{Kruijer17,
       author = {{Kruijer}, Thomas S. and {Burkhardt}, Christoph and {Budde}, Gerrit and {Kleine}, Thorsten},
        title = "{Age of Jupiter inferred from the distinct genetics and formation times of meteorites}",
      journal = {Proceedings of the National Academy of Science},
         year = 2017,
        month = jun,
       volume = {114},
       number = {26},
        pages = {6712-6716},
          doi = {10.1073/pnas.1704461114},
       adsurl = {https://ui.adsabs.harvard.edu/abs/2017PNAS..114.6712K}
}

@INPROCEEDINGS{Bae23,
       author = {{Bae}, J. and {Isella}, A. and {Zhu}, Z. and {Martin}, R. and {Okuzumi}, S. and {Suriano}, S.},
        title = "{Structured Distributions of Gas and Solids in Protoplanetary Disks}",
    booktitle = {Protostars and Planets VII},
         year = 2023,
       editor = {{Inutsuka}, S. and {Aikawa}, Y. and {Muto}, T. and {Tomida}, K. and {Tamura}, M.},
       series = {Astronomical Society of the Pacific Conference Series},
       volume = {534},
        month = jul,
        pages = {423},
          doi = {10.48550/arXiv.2210.13314},
archivePrefix = {arXiv},
       eprint = {2210.13314},
 primaryClass = {astro-ph.EP},
       adsurl = {https://ui.adsabs.harvard.edu/abs/2023ASPC..534..423B}
}

@ARTICLE{Andrews20,
       author = {{Andrews}, Sean M.},
        title = "{Observations of Protoplanetary Disk Structures}",
      journal = {\araa},
         year = 2020,
        month = aug,
       volume = {58},
        pages = {483-528},
          doi = {10.1146/annurev-astro-031220-010302},
archivePrefix = {arXiv},
       eprint = {2001.05007},
 primaryClass = {astro-ph.EP},
       adsurl = {https://ui.adsabs.harvard.edu/abs/2020ARA&A..58..483A}
}

@ARTICLE{KanagawaTanaka17,
       author = {{Kanagawa}, Kazuhiro D. and {Tanaka}, Hidekazu and {Muto}, Takayuki and {Tanigawa}, Takayuki},
        title = "{Modelling of deep gaps created by giant planets in protoplanetary disks}",
      journal = {\pasj},
         year = 2017,
        month = dec,
       volume = {69},
       number = {6},
          eid = {97},
        pages = {97},
          doi = {10.1093/pasj/psx114},
archivePrefix = {arXiv},
       eprint = {1609.02706},
 primaryClass = {astro-ph.EP},
       adsurl = {https://ui.adsabs.harvard.edu/abs/2017PASJ...69...97K}
}

@ARTICLE{KanagawaTanaka15,
       author = {{Kanagawa}, K.~D. and {Tanaka}, H. and {Muto}, T. and {Tanigawa}, T. and {Takeuchi}, T.},
        title = "{Formation of a disc gap induced by a planet: effect of the deviation from Keplerian disc rotation}",
      journal = {\mnras},
         year = 2015,
        month = mar,
       volume = {448},
       number = {1},
        pages = {994-1006},
          doi = {10.1093/mnras/stv025},
archivePrefix = {arXiv},
       eprint = {1501.05422},
 primaryClass = {astro-ph.EP},
       adsurl = {https://ui.adsabs.harvard.edu/abs/2015MNRAS.448..994K}
}

@ARTICLE{Kobayashi12,
       author = {{Kobayashi}, Hiroshi and {Ormel}, Chris W. and {Ida}, Shigeru},
        title = "{Rapid Formation of Saturn after Jupiter Completion}",
      journal = {\apj},
         year = 2012,
        month = sep,
       volume = {756},
       number = {1},
          eid = {70},
        pages = {70},
          doi = {10.1088/0004-637X/756/1/70},
archivePrefix = {arXiv},
       eprint = {1207.1935},
 primaryClass = {astro-ph.EP},
       adsurl = {https://ui.adsabs.harvard.edu/abs/2012ApJ...756...70K}
}

@ARTICLE{Chatterjee14,
       author = {{Chatterjee}, Sourav and {Tan}, Jonathan C.},
        title = "{Inside-out Planet Formation}",
      journal = {\apj},
         year = 2014,
        month = jan,
       volume = {780},
       number = {1},
          eid = {53},
        pages = {53},
          doi = {10.1088/0004-637X/780/1/53},
archivePrefix = {arXiv},
       eprint = {1306.0576},
 primaryClass = {astro-ph.EP},
       adsurl = {https://ui.adsabs.harvard.edu/abs/2014ApJ...780...53C}
}

@ARTICLE{Ayliffe12,
       author = {{Ayliffe}, Ben A. and {Laibe}, Guillaume and {Price}, Daniel J. and {Bate}, Matthew R.},
        title = "{On the accumulation of planetesimals near disc gaps created by protoplanets}",
      journal = {\mnras},
         year = 2012,
        month = jun,
       volume = {423},
       number = {2},
        pages = {1450-1462},
          doi = {10.1111/j.1365-2966.2012.20967.x},
archivePrefix = {arXiv},
       eprint = {1203.4953},
 primaryClass = {astro-ph.EP},
       adsurl = {https://ui.adsabs.harvard.edu/abs/2012MNRAS.423.1450A}
}

@ARTICLE{Lyra09,
       author = {{Lyra}, W. and {Johansen}, A. and {Klahr}, H. and {Piskunov}, N.},
        title = "{Standing on the shoulders of giants. Trojan Earths and vortex trapping in low mass self-gravitating protoplanetary disks of gas and solids}",
      journal = {\aap},
         year = 2009,
        month = jan,
       volume = {493},
       number = {3},
        pages = {1125-1139},
          doi = {10.1051/0004-6361:200810797},
archivePrefix = {arXiv},
       eprint = {0810.3192},
 primaryClass = {astro-ph},
       adsurl = {https://ui.adsabs.harvard.edu/abs/2009A&A...493.1125L}
}

@ARTICLE{Zhu14,
       author = {{Zhu}, Zhaohuan and {Stone}, James M. and {Rafikov}, Roman R. and {Bai}, Xue-ning},
        title = "{Particle Concentration at Planet-induced Gap Edges and Vortices. I. Inviscid Three-dimensional Hydro Disks}",
      journal = {\apj},
         year = 2014,
        month = apr,
       volume = {785},
       number = {2},
          eid = {122},
        pages = {122},
          doi = {10.1088/0004-637X/785/2/122},
archivePrefix = {arXiv},
       eprint = {1308.0648},
 primaryClass = {astro-ph.EP},
       adsurl = {https://ui.adsabs.harvard.edu/abs/2014ApJ...785..122Z}
}

@ARTICLE{Stammler19,
       author = {{Stammler}, Sebastian M. and {Dr{\k{a}}{\.z}kowska}, Joanna and {Birnstiel}, Til and {Klahr}, Hubert and {Dullemond}, Cornelis P. and {Andrews}, Sean M.},
        title = "{The DSHARP Rings: Evidence of Ongoing Planetesimal Formation?}",
      journal = {\apjl},
         year = 2019,
        month = oct,
       volume = {884},
       number = {1},
          eid = {L5},
        pages = {L5},
          doi = {10.3847/2041-8213/ab4423},
archivePrefix = {arXiv},
       eprint = {1909.04674},
 primaryClass = {astro-ph.EP},
       adsurl = {https://ui.adsabs.harvard.edu/abs/2019ApJ...884L...5S}
}

@ARTICLE{Kalyaan21,
       author = {{Kalyaan}, Anusha and {Pinilla}, Paola and {Krijt}, Sebastiaan and {Mulders}, Gijs D. and {Banzatti}, Andrea},
        title = "{Linking Outer Disk Pebble Dynamics and Gaps to Inner Disk Water Enrichment}",
      journal = {\apj},
         year = 2021,
        month = nov,
       volume = {921},
       number = {1},
          eid = {84},
        pages = {84},
          doi = {10.3847/1538-4357/ac1e96},
archivePrefix = {arXiv},
       eprint = {2109.02687},
 primaryClass = {astro-ph.EP},
       adsurl = {https://ui.adsabs.harvard.edu/abs/2021ApJ...921...84K}
}

@ARTICLE{Rice06,
       author = {{Rice}, W.~K.~M. and {Armitage}, Philip J. and {Wood}, Kenneth and {Lodato}, G.},
        title = "{Dust filtration at gap edges: implications for the spectral energy distributions of discs with embedded planets}",
      journal = {\mnras},
         year = 2006,
        month = dec,
       volume = {373},
       number = {4},
        pages = {1619-1626},
          doi = {10.1111/j.1365-2966.2006.11113.x},
archivePrefix = {arXiv},
       eprint = {astro-ph/0609808},
 primaryClass = {astro-ph},
       adsurl = {https://ui.adsabs.harvard.edu/abs/2006MNRAS.373.1619R}
}

@ARTICLE{Morbidelli16,
       author = {{Morbidelli}, A. and {Bitsch}, B. and {Crida}, A. and {Gounelle}, M. and {Guillot}, T. and {Jacobson}, S. and {Johansen}, A. and {Lambrechts}, M. and {Lega}, E.},
        title = "{Fossilized condensation lines in the Solar System protoplanetary disk}",
      journal = {\icarus},
         year = 2016,
        month = mar,
       volume = {267},
        pages = {368-376},
          doi = {10.1016/j.icarus.2015.11.027},
archivePrefix = {arXiv},
       eprint = {1511.06556},
 primaryClass = {astro-ph.EP},
       adsurl = {https://ui.adsabs.harvard.edu/abs/2016Icar..267..368M}
}

@ARTICLE{Desch18,
       author = {{Desch}, Steven J. and {Kalyaan}, Anusha and {O'D. Alexander}, Conel M.},
        title = "{The Effect of Jupiter's Formation on the Distribution of Refractory Elements and Inclusions in Meteorites}",
      journal = {\apjs},
         year = 2018,
        month = sep,
       volume = {238},
       number = {1},
          eid = {11},
        pages = {11},
          doi = {10.3847/1538-4365/aad95f},
archivePrefix = {arXiv},
       eprint = {1710.03809},
 primaryClass = {astro-ph.EP},
       adsurl = {https://ui.adsabs.harvard.edu/abs/2018ApJS..238...11D}
}

@ARTICLE{Hirose11,
   author = {{Hirose}, S. and {Turner}, N.~J.},
    title = "{Heating and Cooling Protostellar Disks}",
  journal = {\apjl},
archivePrefix = "arXiv",
   eprint = {1104.0004},
 primaryClass = "astro-ph.EP",
     year = 2011,
    month = may,
   volume = 732,
      eid = {L30},
    pages = {L30},
      doi = {10.1088/2041-8205/732/2/L30},
   adsurl = {http://adsabs.harvard.edu/abs/2011ApJ...732L..30H}
}

@ARTICLE{DongZhuRafikov15,
       author = {{Dong}, Ruobing and {Zhu}, Zhaohuan and {Rafikov}, Roman R. and {Stone}, James M.},
        title = "{Observational Signatures of Planets in Protoplanetary Disks: Spiral Arms Observed in Scattered Light Imaging Can be Induced by Planets}",
      journal = {\apjl},
         year = 2015,
        month = aug,
       volume = {809},
       number = {1},
          eid = {L5},
        pages = {L5},
          doi = {10.1088/2041-8205/809/1/L5},
archivePrefix = {arXiv},
       eprint = {1507.03596},
 primaryClass = {astro-ph.EP},
       adsurl = {https://ui.adsabs.harvard.edu/abs/2015ApJ...809L...5D}
}

@ARTICLE{KanagawaMuto15,
   author = {{Kanagawa}, K.~D. and {Muto}, T. and {Tanaka}, H. and {Tanigawa}, T. and 
	{Takeuchi}, T. and {Tsukagoshi}, T. and {Momose}, M.},
    title = "{Mass Estimates of a Giant Planet in a Protoplanetary Disk from the Gap Structures}",
  journal = {\apjl},
archivePrefix = "arXiv",
   eprint = {1505.04482},
 primaryClass = "astro-ph.EP",
     year = 2015,
    month = jun,
   volume = 806,
      eid = {L15},
    pages = {L15},
      doi = {10.1088/2041-8205/806/1/L15},
   adsurl = {http://adsabs.harvard.edu/abs/2015ApJ...806L..15K}
}

@ARTICLE{Paardekooper06,
   author = {{Paardekooper}, S.-J. and {Mellema}, G.},
    title = "{Dust flow in gas disks in the presence of embedded planets}",
  journal = {\aap},
   eprint = {astro-ph/0603132},
     year = 2006,
    month = jul,
   volume = 453,
    pages = {1129-1140},
      doi = {10.1051/0004-6361:20054449},
   adsurl = {http://adsabs.harvard.edu/abs/2006A%26A...453.1129P}
}

@INPROCEEDINGS{Whipple72,
   author = {{Whipple}, F.~L.},
    title = "{On certain aerodynamic processes for asteroids and comets}",
booktitle = {From Plasma to Planet},
     year = 1972,
   editor = {{Elvius}, A.},
    pages = {211},
   adsurl = {http://adsabs.harvard.edu/abs/1972fpp..conf..211W},
  publisher = {Wiley},
  address = {New York}
}

@ARTICLE{Eriksson20,
       author = {{Eriksson}, Linn E.~J. and {Johansen}, Anders and {Liu}, Beibei},
        title = "{Pebble drift and planetesimal formation in protoplanetary discs with embedded planets}",
      journal = {\aap},
         year = 2020,
        month = mar,
       volume = {635},
          eid = {A110},
        pages = {A110},
          doi = {10.1051/0004-6361/201937037},
archivePrefix = {arXiv},
       eprint = {2001.11042},
 primaryClass = {astro-ph.EP},
       adsurl = {https://ui.adsabs.harvard.edu/abs/2020A&A...635A.110E}
}

@ARTICLE{ZhuNelson12,
   author = {{Zhu}, Z. and {Nelson}, R.~P. and {Dong}, R. and {Espaillat}, C. and 
	{Hartmann}, L.},
    title = "{Dust Filtration by Planet-induced Gap Edges: Implications for Transitional Disks}",
  journal = {\apj},
archivePrefix = "arXiv",
   eprint = {1205.5042},
 primaryClass = "astro-ph.SR",
     year = 2012,
    month = aug,
   volume = 755,
      eid = {6},
    pages = {6},
      doi = {10.1088/0004-637X/755/1/6},
   adsurl = {http://adsabs.harvard.edu/abs/2012ApJ...755....6Z}
}

@ARTICLE{PinillaBenisty12,
   author = {{Pinilla}, P. and {Benisty}, M. and {Birnstiel}, T.},
    title = "{Ring shaped dust accumulation in transition disks}",
  journal = {\aap},
archivePrefix = "arXiv",
   eprint = {1207.6485},
 primaryClass = "astro-ph.EP",
     year = 2012,
    month = sep,
   volume = 545,
      eid = {A81},
    pages = {A81},
      doi = {10.1051/0004-6361/201219315},
   adsurl = {http://adsabs.harvard.edu/abs/2012A%26A...545A..81P}
}

@ARTICLE{Lubow06,
       author = {{Lubow}, S.~H. and {D'Angelo}, G.},
        title = "{Gas Flow across Gaps in Protoplanetary Disks}",
      journal = {\apj},
         year = 2006,
        month = apr,
       volume = {641},
       number = {1},
        pages = {526-533},
          doi = {10.1086/500356},
archivePrefix = {arXiv},
       eprint = {astro-ph/0512292},
 primaryClass = {astro-ph},
       adsurl = {https://ui.adsabs.harvard.edu/abs/2006ApJ...641..526L}
}

@ARTICLE{ArzamasskiyRafikov18,
       author = {{Arzamasskiy}, Lev and {Rafikov}, Roman R.},
        title = "{Disk Accretion Driven by Spiral Shocks}",
      journal = {\apj},
         year = 2018,
        month = feb,
       volume = {854},
       number = {2},
          eid = {84},
        pages = {84},
          doi = {10.3847/1538-4357/aaa8e8},
archivePrefix = {arXiv},
       eprint = {1710.01304},
 primaryClass = {astro-ph.EP},
       adsurl = {https://ui.adsabs.harvard.edu/abs/2018ApJ...854...84A}
}

@ARTICLE{Gammie01,
       author = {{Gammie}, Charles F.},
        title = "{Nonlinear Outcome of Gravitational Instability in Cooling, Gaseous Disks}",
      journal = {\apj},
         year = 2001,
        month = may,
       volume = {553},
       number = {1},
        pages = {174-183},
          doi = {10.1086/320631},
archivePrefix = {arXiv},
       eprint = {astro-ph/0101501},
 primaryClass = {astro-ph},
       adsurl = {https://ui.adsabs.harvard.edu/abs/2001ApJ...553..174G}
}

@ARTICLE{Shakura73,
   author = {{Shakura}, N.~I. and {Sunyaev}, R.~A.},
    title = "{Black holes in binary systems. Observational appearance.}",
  journal = {\aap},
     year = 1973,
   volume = 24,
    pages = {337-355},
   adsurl = {http://adsabs.harvard.edu/abs/1973A%26A....24..337S}
}

@ARTICLE{Lynden-Bell74,
   author = {{Lynden-Bell}, D. and {Pringle}, J.~E.},
    title = "{The evolution of viscous discs and the origin of the nebular variables.}",
  journal = {\mnras},
     year = 1974,
    month = sep,
   volume = 168,
    pages = {603-637},
   adsurl = {http://adsabs.harvard.edu/abs/1974MNRAS.168..603L}
}

\appendix

\section{On the choice of the $f_{\rm heat}$ parameter}
\label{sec:fheat}
{
Our model for the steady-state temperature distribution, equation~\eqref{eq:T_predict}, includes an ad-hoc dimensionless parameter $f_{\rm heat}$ accounting for possible non-local cooling effects such as adiabatic expansion and advective heat transport. Here, we calibrate this parameter using our simulation results.

Figure~\ref{fig:fheat} compares the relative temperature enhancement, $T/T_{\rm init}$, obtained from Runs 1--4 with the values predicted by  equation~\eqref{eq:T_predict} using 
$f_{\rm heat} = 1$ and $f_{\rm heat} = 0.8$.
The choice $f_{\rm heat} = 1$ corresponds to the case in which shock heating exactly balances local thermal relaxation. 
We find that this choice overestimates the temperature enhancements around the gap edge ($0.7 \lesssim r/r_{\rm p} \lesssim 1.5$) by up to 15\%. Choosing $f_{\rm heat} = 0.8$ reduces these errors to $\lesssim 4\%$ across all four runs and over the entire computational domain. 
}

\begin{figure*}[t]
\begin{center}
\includegraphics[width=\hsize, bb=0 0  723.125 210]{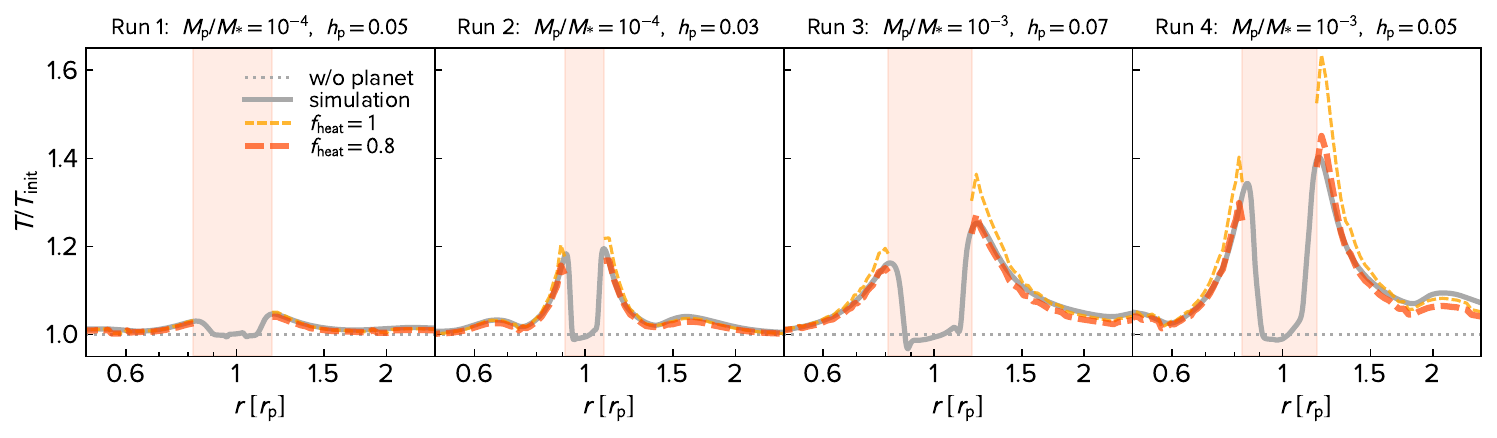}
\end{center}
 \caption{
 Relative temperature enhancement $T/T_{\rm init}$ predicted from semi-analytic models (section~\ref{sec:semi}) with different values of the tuning parameter  $f_{\rm heat}$ in equation~\eqref{eq:T_predict} (dashed lines). Solid lines show the azimuthally averaged profiles for four simulation runs, while thin and thick dashed lines show the model predictions for $f_{\rm heat} =$ 1 and $f_{\rm heat} =$ 0.8 (the value adopted in this study), respectively. Dotted lines represent the profiles before the planet's insertion.  Shaded regions around the planet's orbit are excluded from streamline analysis in the main text. The boundaries of these regions are defined so as to match the locations of the maxima in $T/T_{\rm init}$. 
 }
\label{fig:fheat}
\end{figure*}

\section{Comparison with the empirical gap model by \citet{KanagawaTanaka17}}
\label{sec:Kanagawa}

Here, we compare our simulation results with the empirical gap model proposed by \citet{KanagawaTanaka17}. Their model provides a piecewise expression for the surface density profile around a planet-carved gap, is given by
\begin{equation}
\Sigma(r) = \begin{cases}
\Sigma_{\rm floor}, & |r-r_{\rm p}| < \Delta r_1, \\
\Sigma_{\rm gap}(r), & \Delta r_1 < |r-r_{\rm p}| < \Delta r_2, \\
\Sigma_{\rm init}, & |r-r_{\rm p}| > \Delta r_2,
\end{cases}    
\label{eq:Kanagawa17}
\end{equation}
where $\Sigma_{\rm floor}$ is the floor value of $\Sigma$ at the gap center (see also \citealt{KanagawaMuto15,KanagawaTanaka15}), $\Sigma_{\rm gap}$ represents the surface density profile between the gap center and edge, and 
$\Delta r_1$ and $\Delta r_2$ are the widths of the gap center and whole gap, respectively. These quantities are given by
\begin{equation}
\Sigma_{\rm floor} = \frac{\Sigma_{\rm init}}{1+0.04K},
\end{equation}
\begin{equation}
\Sigma_{\rm gap} = \left(4 K'^{-1/4} \frac{|r-r_{\rm p}|}{r_{\rm p}}-0.32\right)\Sigma_{\rm init},
\end{equation}
\begin{equation}
\Delta r_1 = \left(\frac{\Sigma_{\rm floor}}{4\Sigma_{\rm init}}+0.08\right) K'^{-1/4} r_{\rm p},
\label{eq:Delta_r1}
\end{equation}
\begin{equation}
\Delta r_2 = 0.33 K'^{-1/4} r_{\rm p},
\label{eq:Delta_r2}
\end{equation}
with
\begin{equation}
 K =\frac{1}{\alpha}\pfrac{M_{\rm p}}{M_*}^2 h_{\rm p, K17}^{-5},
 \label{eq:K}
\end{equation}
\begin{equation}
 K' =\frac{1}{\alpha}\pfrac{M_{\rm p}}{M_*}^2 h_{\rm p, K17}^{-3},    
 \label{eq:K'}
\end{equation}
where $h_{\rm p, K17}$ is the ratio between the initial sound speed and Keplerian velocity at $r=r_{\rm p}$ in the isothermal simulations by \citet{KanagawaTanaka17}. Since the disks in our simulations are substantially adiabatic with $\beta = 100$, we interpret $h_{\rm p, K17}$ as the {\it adiabatic} sound speed, i.e., $h_{\rm p, K17} = \sqrt{\gamma}h_{\rm p}$ with $\gamma = 1.4$, when applying the  \citet{KanagawaTanaka17} model to our simulations. 

\begin{figure*}
\begin{center}
\includegraphics[width=\hsize, bb=0 0 727.5 210.068125]{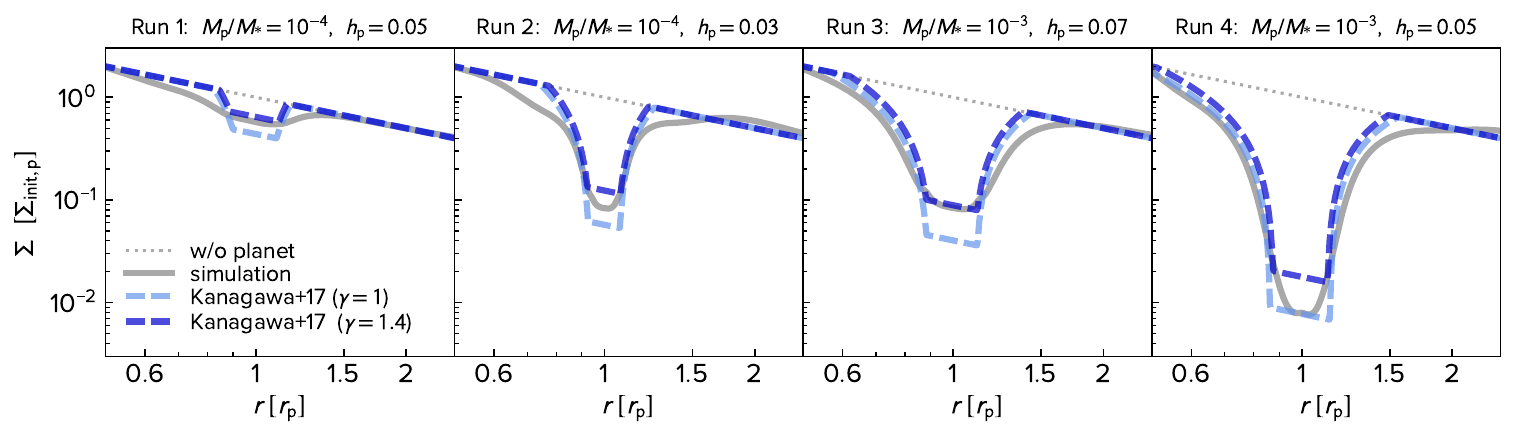}
\end{center}
 \caption{Surface density profiles predicted from the empirical deep-gap model by \citet{KanagawaTanaka17}, both without and with a correction for $h_{\rm p, K17}$ (corresponding to $\gamma =$ 1 and 1.4, respectively; see appendix~\ref{sec:Kanagawa}), compared with the azimuthally averaged surface density profiles from 2D simulations (solid lines).  Dotted lines represent the profiles before the planet's insertion, $\Sigma = \Sigma_{\rm init}$.
 }
\label{fig:Kanagawa}
\end{figure*}

Figure~\ref{fig:Kanagawa} compares the azimuthally averaged surface density from our simulations with the predictions of the \citet{KanagawaTanaka17} model. The correction for $h_{\rm p, K17}$ has a considerable effect on the gap floor value but has a minor impact away from the gap center. Based on their isothermal simulations, \citet{KanagawaTanaka17} reported that their model provides a good estimate for deep gaps but underestimates the widths of shallow ones. Our simulations confirm this, showing that the \citet{KanagawaTanaka17} model reproduces the deep gaps in Runs 3 and 4 more accurately than the shallower gaps in Runs 1 and 2, which are wider than the model predicts. The correction for $h_{\rm p, K17}$ improves the accuracy of the gap floor value; without this correction (i.e., $h_{\rm p, K17} = \sqrt{\gamma}h_{\rm p}$ with $\gamma = 1$), the \citet{KanagawaTanaka17} model consistently underestimates the floor values by a factor of $\gtrsim 1.5$, as shown in figure~\ref{fig:Kanagawa}. 

\begin{figure*}
\begin{center}
\includegraphics[width=\hsize, bb=0 0 727.5 210.068125]{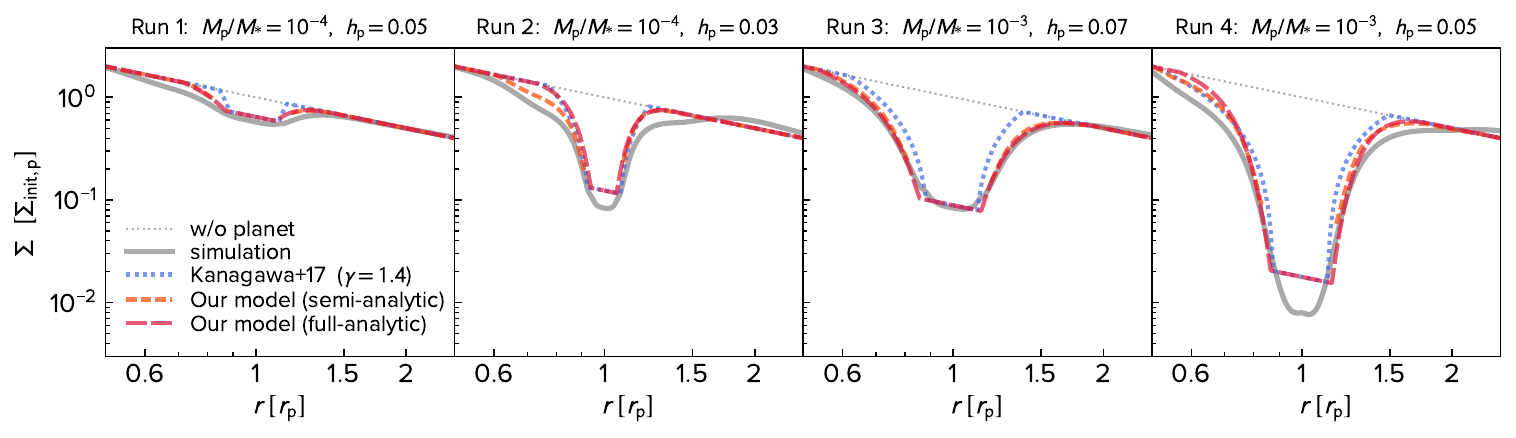}
\end{center}
 \caption{Comparison of radial surface density profiles predicted by different models. Solid lines show azimuthally averaged surface density profiles from four 2D simulations. Thick dotted, dashed, and long-dashed lines represent predictions from the empirical gap model by \citet{KanagawaTanaka17} with a correction for $h_{\rm p, K17}$ (appendix~\ref{sec:Kanagawa}), the semi-analytic model (section~\ref{sec:semi}), and the full-analytic model (section~\ref{sec:analytic}), respectively. Thin dotted lines show the profiles before the planet's insertion, $\Sigma = \Sigma_{\rm init}$.
 }
\label{fig:Kanagawa_comp}
\end{figure*}
In figure~\ref{fig:Kanagawa_comp}, we also compare our simulation results and model predictions with the surface density profiles predicted by the empirical deep-gap model of \citet{KanagawaTanaka17}. For Runs 2 and 4, both our full-analytic model and that of \citet{KanagawaTanaka17} provide comparably good predictions for the surface density profile. In contrast, for Runs 1 and 3, our model more accurately reproduces the widths of the gaps. The discrepancy between the simulation and \citet{KanagawaTanaka17} model is particularly noticeable for the shallow gap observed in Run 1. As noted by \citet{KanagawaTanaka17}, their model does not yield reliable estimates for shallow gaps with $\Sigma_{\rm floor}/\Sigma_{\rm init} \gtrsim 0.5$. 

Finally, we emphasize that the {primary advantage of our gap model over that of \citet{KanagawaTanaka17} lies in its ability to self-consistently predict surface density and temperature profiles for $\beta \gg 1$, rather than in improved accuracy}.

\end{document}